\documentclass[useAMS,usenatbib]{mnras}     
\usepackage{times}

\usepackage{./aas_macros}
\usepackage{graphicx}
\graphicspath{{images/}}
\usepackage{amssymb}
\usepackage{amsmath}
\usepackage{multirow}
\usepackage{color}
\usepackage[normalem]{ulem}

\begin{document}

\title{ Coherent magneto-elastic oscillations in superfluid magnetars}
\author[Michael Gabler, Pablo Cerd\'a-Dur\'an, Nikolaos Stergioulas, Jos\'e
A.~Font and Ewald M\"uller]
{Michael Gabler$^{1}$, 
Pablo Cerd\'a-Dur\'an$^2$, 
Nikolaos Stergioulas$^3$, 
Jos\'e A.~Font$^{2,4}$, 
\and and 
Ewald M\"uller$^1$ 
\\
  $^1$Max-Planck-Institut f\"ur Astrophysik,
  Karl-Schwarzschild-Str.~1, 85741 Garching, Germany \\
  $^2$Departamento de Astronom\'{\i}a y Astrof\'{\i}sica,
  Universidad de Valencia, 46100 Burjassot (Valencia), Spain\\
  $^3$Department of Physics, Aristotle University of Thessaloniki,
  Thessaloniki 54124, Greece \\
  $^4$ Observatori Astron\`omic, Universitat de Val\`encia, C/ Catedr\'atico 
  Jos\'e Beltr\'an 2, 46980, Paterna (Val\`encia), Spain \\
  }
\date{\today}
\maketitle
\begin{abstract}
We study the effect of superfluidity on torsional oscillations of highly 
magnetised neutron stars (magnetars) with a microphysical equation of 
state by means of two-dimensional, magneto-hydrodynamical-elastic simulations. 
The superfluid properties of the neutrons in the neutron star core are treated 
in a parametric way in which we effectively decouple part of the core matter 
from the oscillations. Our simulations confirm the existence of two groups of 
oscillations, namely continuum oscillations that are 
confined to the neutron star core and are of Alfv\'enic character, and global 
oscillations with constant phase and that are of mixed magneto-elastic 
type. The latter might explain 
the quasi-periodic oscillations observed in magnetar giant flares, since they 
do not suffer from the additional damping mechanism due to phase mixing, 
contrary to what happens for continuum oscillations. However, we cannot 
prove rigorously that the coherent oscillations with constant phase are normal 
modes. Moreover, we find no crustal shear modes for the magnetic field 
strengths typical for magnetars. We provide fits to our numerical simulations 
that give the oscillation frequencies as functions of magnetic field strength 
and proton fraction in the core. 
\end{abstract}
%
\section{Introduction}
The analysis of the X-ray light curves of the giant flares of the soft 
gamma-ray repeaters (SGRs) SGR 1806-20 and SGR 1900+14 revealed a number of 
quasi-periodic oscillation (QPOs) with frequencies at $18$, $26$, $30$, 
$92$, $150$, $625$, $1840\,$Hz (SGR 1806-20) and $28$, $53$, $84$, $155$\,Hz 
(SGR 1900+14) \citep[see e.g.][]{Israel2005, 
Strohmayer2005, Watts2006,Strohmayer2006}.  
The importance of this discovery became apparent when the observed QPOs were 
identified as oscillations of the neutron star. If such interpretation were true, 
it would provide a unique tool to study the interior of a neutron star through 
asteroseismology. Since the oscillation frequencies of a neutron star 
depend sensitively on the equation of state (EoS) the matching of a 
theoretical model with the observations would allow to constrain the EoS.
More recent studies found additional QPOs in the flare of SGR 1806-20 
\citep[at $17$, $21$, $36$, $59$, and $116\,$Hz in][]{Hambaryan2011} and also in 
less energetic bursts of SGR 1806-20 \citep[at $57\,$Hz in][]{Huppenkothen2014b} and in SGR J1550-5418 \citep[at $93$, $127$, and $260\,$Hz in][]{Huppenkothen2014a}.
The observed frequencies can thus be neatly divided into two groups, below $260\,$Hz 
and above $500\,$Hz. 

Initially the observed QPO frequencies were 
assumed to be discrete torsional shear oscillations of the 
elastic crust of 
the neutron star. The low-frequency QPOs match approximately the 
frequencies of fundamental torsional crustal shear modes $t_n^l$, i.e. modes 
without nodes 
in the radial direction ($n=0$), while the higher frequencies could correspond to 
overtones ($n>0$). The model of shear modes has been investigated 
carefully by many groups and different aspects of the nuclear 
physics of the crust have been discussed in detail \citep[see][
and references therein]{Duncan1998, Messios2001, Strohmayer2005, Piro2005, Sotani2007,
Samuelsson2007, Steiner2009, Deibel2014, Sotani2016}. 

However, the observed properties of SGRs indicate that the problem is more 
difficult. Their X-ray luminosity, X-ray spectra, spin-down 
measurements and bursting activity together with other observational 
indications strongly suggest that SGRs are highly magnetised neutron stars 
(magnetars) with surface field strengths of up to 
$B\gtrsim10^{15}\,$G \citep{Duncan1992}. These strong magnetic fields lead to 
torsional Alfv\'en oscillations with frequencies around $30\,$Hz 
\citep{Sotani2008,Cerda2009,Colaiuda2009}, exactly in the right frequency 
range of the observed QPO frequencies. The Alfv\'en oscillations do not 
form an eigenmode system but can be found as long-lived QPOs at the turning 
points and edges of an Alfv\'en ontinuum \citep{Levin2007}. 
For simplified models, it was shown by \cite{Levin2006} and 
\cite{Glampedakis2006b} that the interaction of the discrete crustal modes with 
the Alfv\'en continuum leads to strong absorption of the shear modes into the 
core. More realistic calculations confirmed this result and also 
showed that, instead of the crustal modes, 
magneto-elastic oscillations of predominantly Alfv\'enic character have 
frequencies that match the observed low-frequency QPOs.
However, within this magneto-elastic model it is not easy to accommodate the 
highest observed QPO frequencies of $625$ and $1840\,$Hz (SGR 1806-20) 
\citep{vanHoven2011, Gabler2011letter, Colaiuda2011, vanHoven2012, Gabler2012}. 
These would have to be fairly high overtones of the fundamental magneto-elastic 
oscillations, which would render problematic to explain why only such particular 
high-frequency QPOs are excited.  In a more sophisticated analysis of 
the observations, \cite{Huppenkothen2014c} find that the data 
 is consistent with the $625\,$Hz QPO being transient, which indicates a 
strong crust-core coupling.

Most of the above models considered a single fluid in the core of the
neutron star, consisting of a mixture of neutrons, protons and electrons. 
This is a valid approach if the interactions between different
species, such as neutrons and protons, are very strong. However, 
\cite{Migdal1959} and \cite{Baym1969}
discussed the possibility that the neutrons in the core of neutron 
stars are superfluid. This idea received strong support by \cite{Shternin2011} 
and \cite{Page2011} who showed that the cooling curve of Cas A is 
consistent with a phase transition to superfluid neutrons. If the neutrons 
are indeed superfluid, the matter in the core of neutron stars cannot
be described within the single-fluid approach. First studies considering 
the 
effect of superfluid phases on the dynamics of the fluid mixture inside 
neutron stars \citep{Mendell1991, Mendell1998} were followed by
\cite{Prix2002, Andersson2002, Andersson2004, Chamel2008b}. These models 
provided the framework to obtain first estimates of the effects of superfluidity on the
oscillation spectra of magnetised neutron stars {\it without} crust
\citep{Glampedakis2011a, Passamonti2013}. In particular, \cite{Passamonti2013} 
tried to explain 
the high-frequency QPOs as Alfv\'en oscillations but had difficulties in 
identifying the low-frequency QPOs. Additionally, \cite{Samuelsson2009} and 
\cite{Sotani2013} investigated the effect of superfluid neutrons in the crust on the
shear oscillations of the latter, but did not take magnetic fields into 
account. 

The first magneto-elastic models considering superfluid effects were presented in 
\cite{Levin2007, vanHoven2008, vanHoven2011, vanHoven2012}. These 
authors did not find significant 
differences from models without superfluidity other than a shift of the 
fundamental magneto-elastic frequency for a given magnetic field strength to 
higher values, however, which is important to bring the estimated 
magnetic field strength in agreement to those observed from spin-down 
measurements $\bar B\sim 10^{15}\,$G \cite{vanHoven2008}. Subsequent work by 
\cite{Gabler2013b} revealed that the
inclusion of superfluid effects does not only modify the frequencies (such 
that the spin-down estimates for the global magnetic field strength are in 
better agreement with QPOs originating from magneto-elastic oscillations), but 
also that there 
exists a different type of QPOs that exhibit a \textit{constant phase} 
throughout the star. Additionally, \cite{Gabler2013b} 
discovered high frequency oscillations that may explain the observed QPOs at 
$625$ and $1840\,$Hz. These results were confirmed in a study by 
\cite{Passamonti2014}. The evolutionary paths of the different (superfluid)
magneto-elastic oscillations in magnetars were also analysed using 
order-of-magnitude 
estimates of different damping time scales in \cite{Glampedakis2014b}.

In this work we extend the study initiated in \cite{Gabler2013b} and discuss in detail 
the effects of superfluidity on the spectrum of magneto-elastic 
oscillations of magnetars. We first review in Section~\ref{framework} the equations for magneto-elastic oscillations in 
the single-fluid approach and then generalize these equations to the case of 
superfluid neutrons. 
We also update the equilibrium model used in \cite{Gabler2013b} with a 
magnetic field configuration that takes superfluid neutrons into account and 
discuss the boundary conditions for the simulations. In Section 
\ref{sec_low_f}, we discuss the effects of superfluid neutrons on the 
low-frequency magneto-elastic oscillations. We confirm the existence of a new 
class of coherent oscillations with constant phases 
\citep{Gabler2013b,Passamonti2014} 
and are able to distinguish two families: predominantly Alfv\'en QPOs that are 
confined to the core and have a continuous phase, and global magneto-elastic 
oscillations that have a constant phase. Then, we cast our ignorance about the 
superfluid entrainment factor and the proton fraction into one effective 
parameter and study the effects of a change of the latter.
Finally, we discuss our results in Section\,\ref{sec_discuss}.

Throughout this work we will use units where $c=G=1$,  where, $c$ and 
$G$ are the speed of light and the gravitational constant, respectively. Latin 
(Greek) indices run from 1 to 3 (0 to 3). Partial derivatives are 
indicated by a comma and we apply the Einstein summation convention.
The magnetic field strength we use is that of a uniformly magnetised rotating 
sphere with a radius of $10\,$km that would cause the same magnetic dipole 
spin-down as our stellar model (see Section\,\ref{sec_bfield} for details).

%
\section{Theoretical framework}
\label{framework}

\subsection{Evolution equations}

This study is based on the numerical code {\tt MCOCOA} that solves the
general-relativistic magneto-hydrodynamical (GRMHD) equations and includes a
description for elasticity. The implementation of GRMHD was discussed and
tested carefully in \cite{Cerda2008, Cerda2009} and the treatment of elasticity 
was included in \cite{Gabler2011letter, Gabler2012, Gabler2013a}. We assume a 
spherically symmetric spacetime, keep the space time fixed (Cowling 
approximation), and use the corresponding line element in 
isotropic coordinates
\begin{eqnarray}
 ds^2 = - \alpha^2 dt^2 + \Phi^4\hat\gamma_{ij} dx^i dx^j, \
\end{eqnarray}
where $\alpha$ and $\Phi$ are the lapse function and the conformal factor,
respectively, and $\hat\gamma_{ij}={\rm diag}(1,r^2,r^2 \sin\theta)$ is 
the spatial, flat 3-metric. The 
circumferential radius $R$ is thus related to the coordinate radius $r$ by 
$R=\Phi^2 r$.

The torsional magneto-elastic oscillations follow from the
conservation of energy and momentum, baryon number conservation and
Maxwell's equations. If we further consider only linear perturbations in 
axisymmetry, poloidal 
and toroidal perturbations decouple and we can  
write down the following system of equations \citep{Gabler2011letter, 
Gabler2012}:
\begin{eqnarray}
 \frac{1}{\sqrt{-g}} \left( \frac{\partial\sqrt{\gamma} \mathbf{U}
}{\partial t} +
\frac{\partial \sqrt{-g} \mathbf{F}^i}{\partial x^i} \right) = 0\,,
\label{conservationlaw}
\end{eqnarray}
where $g$ is the determinant of the metric, $\gamma$ is the determinant of the
three-metric and the state vector $\bf U$ and flux vectors $\bf F^i$ are
given by:
\begin{eqnarray}
 \mathbf {U} &=& [S_\varphi, B^\varphi]  \label{reduced_withcrust1}\,,\\
 \mathbf {F}^r &=& \left[ -
\frac{b_\varphi B^r}{W} - 2 \mu_\mathrm{S}
\Sigma^r_{~\varphi}, - v^{\varphi} B^r
\right]\,,  \label{flux_r}\\
 \mathbf{ F}^\theta &=& \left[ - \frac{b_\varphi B^\theta}{W}- 2
\mu_\mathrm{S} \Sigma^\theta_{~\varphi},
-v^{\varphi} B^\theta
\right]\,.\label{flux_theta}\label{reduced_withcrust2}
\end{eqnarray}
Here, $B^i$ is the magnetic field measured by an Eulerian observer, $b_i$
that of a co-moving observer, $W=\alpha u^{t}$  the Lorentz factor,
$v^{i}$ the three-velocity, $\mu_\mathrm{S}$ the shear modulus,
$\Sigma^{\mu\nu}$ the shear tensor, and
$S_i=(\rho h + b^2) W^{2} v_i - \alpha b_i b^0$ is a
generalisation of the momentum density, with $\rho$ and $h$ being the rest-mass
density and specific enthalpy, respectively.

To solve these equations numerically, we need to know the two components of the
shear tensor $\Sigma^r_{~\varphi}=g_{\varphi\varphi} g^{rr} \xi^\varphi_{,r}$
and $\Sigma^\theta_{~\varphi}=g_{\varphi\varphi} g^{\theta\theta}
\xi^\varphi_{,\theta}$ that depend on the spatial derivatives of the
displacement $\xi^\varphi$. The latter is related to the three velocity by
\begin{eqnarray}
\xi^j_{\,,t} = \alpha v^j \label{def_xidot}\,.
\end{eqnarray}
Because the order of the partial derivatives can be interchanged, we use
the following equations for the spatial derivatives of $\xi^\varphi$
\begin{eqnarray} 
(\xi^\varphi_{\,,r})_{,t} - (v^\varphi \alpha)_{,r} &=&0\,,\label{eq_xi_dr}\\
(\xi^\varphi_{\,,\theta})_{,t} - (v^\varphi \alpha)_{,\theta} &=&
0\label{eq_xi_dtheta}\,.
\end{eqnarray}

The complete system of equations of elastic GRMHD to be solved 
is given by Eqs.\,(\ref{conservationlaw}), (\ref{eq_xi_dr}) and (\ref{eq_xi_dtheta}) 
together with the corresponding definitions in Eqs.\, 
(\ref{reduced_withcrust1})--(\ref{reduced_withcrust2}). This set of equations 
assumes a single constituent fluid, i.e. all particles in the core respond 
together to any perturbation. However, theoretical \citep{Baym1969} and
observational \citep{Anderson1975, Shternin2011, Page2011} results indicate that
neutrons are superfluid and protons are superconducting in old 
($t\gtrsim100$y) neutron star cores. In this work we consider magnetars with 
sufficiently strong magnetic fields ({ $B\gtrsim \mathrm{few} \times 
10^{15}\,$G at the surface and $B\sim10^{16}\,$G in the core}) 
to suppress the proton 
superconductivity in their cores \citep{Glampedakis2011a, Sinha2015}. As a result, 
and also because the MHD equations for superconducting protons 
 \cite[presently being developed by][]{Graber2015} differ significantly from those of standard MHD, we
neglect the effects of superconductivity and use the standard MHD 
equations. 

For our neutron star models we assume a normal-matter EoS, i.e. 
we allow for neutrons, protons, electrons and muons. The latter three 
species are coupled together by the electromagnetic force. For the dynamics of 
this conglomerate the mass of the electrons (and muons) is negligible
compared to that of the protons. Thus, we refer to all other constituents but
superfluid neutrons as charged particles. In the crust the charged nuclei of 
the crustal lattice dominate the evolution instead of the protons. For
the time scales considered here, the electrons and muons are assumed to co-move 
with the nuclei. Between the superfluid neutrons and the charged particles
there is no direct interaction, i.e. the hydrodynamical equations of both fluids
decouple. They interact with each other only through the gravitational
potential. In contrast, in a mixture of superfluid neutrons and
superconducting charged particles (protons) in the core, both
superfluid components feel each other through the strong interaction and each
of the superfluid species `entrains' part of the other. For non-superconducting
protons the corresponding entrainment coefficients $\varepsilon_p$ and
$\varepsilon_n$ are zero \citep{Gusakov2005}, where $n$ and $p$ denote
neutrons and protons, respectively. 
The entrainment is related to the concept of effective mass $m^\ast$
\begin{eqnarray}
 \varepsilon_x = 1 - \frac{m^\ast_x}{m_x}\,,
\end{eqnarray}
where $m_x$ is the bare mass of the species $x\in\{n,~p\}$. 

In Newtonian models, it was shown by \cite{Andersson2009} that the momentum
equation of the neutrons for an incompressible fluid with constant 
background (which is equivalent to the assumption made for torsional 
oscillations) is coupled to the protons by a direct proportionality between the
proton and neutron accelerations \citep[see Eq.~(18) in][]{Andersson2009}.
The proportionality factor is related to the entrainment parameter
of the neutrons $\varepsilon_n / (1-\varepsilon_n)$. This proportionality can
be substituted into the equation of motion of the charged particles where the 
neutron acceleration appears. There are two resulting differences in the 
equations for the charged particles compared
to the single-fluid approach. First, the density in the momentum equations that
was the total fluid density in non-superfluid models has to be replaced by the
density of the charged particles only, i.e. $\rho \rightarrow \rho_c = X_c \rho$,
where $X_c$ is the mass fraction of charged particles. Second, in all
places where the density appears, a factor related to the entrainment parameters has 
to be included:
\begin{eqnarray}
\varepsilon_\star = \frac{1-\varepsilon_n}{1-\varepsilon_n-\varepsilon_c}\,.
\end{eqnarray}

Although not taking the superconductive properties of the protons explicitly
into account for the evolution equations, we include the entrainment in our
calculations. Since we know neither the
exact EoS of the core, nor the fraction of charged particles or the entrainment
between different particles, we incorporate all these uncertainties in
a parametrization of the combination $\varepsilon_\star X_c$, which always 
will appear together. This will also allow for the possibility that not the 
entire 
core is in the superfluid state. The combination $\varepsilon_\star X_c$ 
describes how much of the core matter is taking part in the Alfv\'en motion. If 
$\varepsilon_\star X_c=1,$ the equations do not change and all matter behaves as 
in the single fluid model.

The qualitative results we obtain with our normal MHD model may also be valid
for superconducting protons. We know that the magnitude of the local speed of
propagation of Newtonian Alfv\'en waves, given by $v_A^2=B^2/\rho_c$, is similar to that of
their superconductive counterparts, the Cyclotron-Vortex waves $v^2=B 
H_{c1} / \rho_c$. Here, $H_{c1}\sim10^{15}\,$G is the first critical magnetic 
field of the II superconductor that is expected in neutron star cores. 
While we expect some quantitative changes of the results, the general picture we 
present here with the current method should also be valid for superconducting 
protons. The question of superconductivity will be addressed in further studies.

Superfluid neutrons in a rotating star will form neutron vortices. These 
vortices may couple to the crust by pinning and thus may also be excited by 
torsional oscillations. However, magnetars are rotating so slowly that only a 
tiny fraction of neutrons forms vortices. The neutron vortices may 
also be pinned to the proton-electron plasma in the core, which forms fluxtubes 
if the protons are superconducting. The latter pinning was shown to be 
inefficient by \cite{vanHoven2008} and we can safely neglect pinning effects 
for our problem of magneto-elastic oscillations. The possibly entrained 
protons that may magnetise the vortices are also negligible. 
Actually, the additional particles participating in the torsional oscillations 
are already included in our parametrization of $\varepsilon_\star X_c$ and 
would increase the proton fraction only slightly.

In our \textit{effective single-fluid approximation} to the superfluid model,
Eqs.~(\ref{reduced_withcrust1})--(\ref{reduced_withcrust2}) of the 
non-superfluid case are changed to
\begin{eqnarray}
 \mathbf{U} &=& [S^{(c)}_\varphi, B^\varphi]  \label{reduced_withcrust1_c}\,,\\
 \mathbf{F}^r &=& \left[ -
\frac{b_\varphi B^r}{W^{(c)}} - 2 \mu_\mathrm{S}
\Sigma^r_{~\varphi}, - v^{\varphi(c)} B^r
\right]\,,  \label{flux_r_p}\\
 \mathbf{F}^\theta &=& \left[ - \frac{b_\varphi B^\theta}{W^{(c)}}- 2
\mu_\mathrm{S} \Sigma^\theta_{~\varphi},
-v^{\varphi(c)} B^\theta
\right]\,,\label{flux_theta_p}\label{reduced_withcrust2_c}
\end{eqnarray}
where we introduce $W^{(c)}=\alpha u^{t(c)}$ the Lorentz factor,
$v^{i(c)}$ the three-velocity, and $S^{(c)}_i=(\varepsilon_\star X_c \rho h +
b^2) W^{(c)2} v^{(c)}_i - \alpha b_i b^0$ the momentum density of the {\it charged}
particles. Note that the entrainment factor $\varepsilon_\star$ and the mass
fraction of charged particles $X_c$ are responsible for a
qualitatively different behaviour compared to the normal fluid approach.
Eqs.\,(\ref{eq_xi_dr}) and (\ref{eq_xi_dtheta}) are generalised by simply substituting
$\xi^\varphi\rightarrow \xi^{\varphi(c)}$ and $v^\varphi \rightarrow
v^{\varphi(c)}$.

The eigenvalue problem of the flux-vector Jacobian associated 
with Eqs.~(\ref{reduced_withcrust1_c})--(\ref{reduced_withcrust2_c})
has the following non-zero eigenvalues
\begin{eqnarray}
\lambda_{1/2}^{k} &=& \pm \sqrt{\frac{ (B^k)^2 + \mu_\mathrm{S} /
g_{kk} }{A}}\,, \label{eq:eigenvalues} ~~~k=\{r,\theta\}\label{eq_EV_1}\,
\end{eqnarray}
with
\begin{eqnarray}
A = \frac{\partial S^{(c)}_\varphi}{\partial v_\varphi^{(c)}} = 
\varepsilon_\star 
X_c\rho h W^{(c)4} \left(1+v_\varphi^{(c)}
v^{\varphi(c)}\right) 
+ B^k B_k \,.\label{eq_EV_2}
\end{eqnarray}

A similar approach was also followed by \cite{Passamonti2013} in their Newtonian
model. However, they considered only stratified polytropes with prescribed
entrainment. We are not aware of any computational tool to study
the general case of arbitrary coupling/entrainment between neutrons and protons
in general relativity, including magnetic fields and the solid crust. 

\subsection{Magnetar equilibrium model}

We want to study the QPOs of magnetars that have frequencies $f\gtrsim20\,$Hz.
The dynamical time scale of interest is thus several oscillations periods
$t_\mathrm{QPO}\lesssim1\,$s. Since this is much shorter than a typical 
rotation period of a magnetar, $t_\mathrm{rot}\sim 2-12\,$s, neglecting
rotation is a very good approximation. A non-rotating, 
unmagnetised equilibrium model is the same for superfluid or normal neutrons in 
the core of the neutron star. The different constituents feel each other 
exclusively through their gravitational interaction that is the same in both 
cases. However, magnetic fields are essential. First magnetised models of  neutron stars with superfluid neutrons in the core have been obtained for polytropic EoS by \cite{Passamonti2013},
while   \cite{Palapanidis2015} investigated the effect of entrainment if the neutrons are considered to be superconducting (see also references therein).

We construct the stratified fluid equilibrium model with a version of the RNS
code \citep{Stergioulas1995} that was extended to solve for the dipole magnetic 
field structure in the approximation of a passive field (the magnetic field is 
assumed to have no influence the fluid equilibrium, which is an accurate 
description for magnetic field strengths considered here). Hence, as a first 
step, a nonrotating fluid equilibrium is obtained and then, as a second step, 
the MHD equations for a dipolar configuration are solved, following closely the 
formalism by \citep{Bocquet1995}. Superfluid effects are mimicked by 
introducing the proton fraction $X_c$ as a parameter. $X_c$ appears as a 
multiplicative factor at the right side of equation (15) in the original 
approach by 
\cite{Bocquet1995}. 

We choose one sample EoS that is a combination of the
Akmal-Pandharipande-Ravenhall (APR) EoS for the core \citep{Akmal1998} and the
Douchin-Hanesel (DH) EoS for the crust \citep{Douchin2001}. The reference model
has a mass of $1.4\mathrm{M}_\odot$ and a radius of $R=12.26\,$km 
($r=10.08\,$km). \citet{Douchin2001} also directly provide the proton fraction, 
which we use as reference value $X_c^0(r)$.
In the core we take the effective masses $m_n^\ast$ and $m_p^\ast$ that are
provided by \cite{Chamel2008} for a parametrised nuclear force NRAPR that was
presented in \cite{Steiner2005}. With the effective masses we can
calculate the entrainment parameters $\varepsilon_n$ and $\varepsilon_p$
and the reference entrainment factor $\varepsilon^0_\star(r)$. 

In the inner crust at densities above the neutron drip point, superfluid
neutrons should be present, while the nuclei which are organised in a lattice
are not superfluid. However, the neutrons can be scattered by the regular
nuclear lattice \citep{Chamel2012}. This Bragg reflection leads to a very
efficient entrainment that couples a large part of the superfluid neutrons to
the nuclei. The corresponding calculation is not available for the DH EoS.
However, for most parts of the crust the number of conduction neutrons, i.e.
neutrons conducting the superfluid neutron current and therefore decoupling
from the shear and Alfv\'en motion of the lattice nuclei, is roughly 10\%
of the total nucleons \citep{Chamel2012}. We will take this value as an
approximation throughout this work. In the inner crust, we thus set $X_c = 0.9$
and $\varepsilon_\star=1.0$. 

\subsection{Boundary conditions}

The boundary conditions for our simulations have been discussed extensively in 
\cite{Gabler2012}. They are based on the assumptions of MHD 
and conservation of momentum. As a direct consequence, the velocity and 
thus the displacement have to be
continuous everywhere in the star. This holds in particular at the surface and
at the core-crust interface. At the surface we allow for toroidal magnetic
field perturbations that are supported by currents in a twisted, 
force-free magnetosphere, as expected according to
\cite{Thompson2002}. In a force-free configuration, the magnetic field 
and the currents are parallel and proportional to each other. This implies that
the field has to be continuous across the surface \citep[see 
also][for a numerical confirmation]{Gabler2014b}.
The corresponding boundary conditions are thus continuity of the magnetic field 
and continuous traction. These conditions lead to
\begin{eqnarray}
 b^\varphi_\mathrm{crust} &=& b^\varphi_\mathrm{atmosphere}\,,\\
\xi^\varphi_{crust,r}&=&0\,.
\end{eqnarray}

At the core-crust interface the continuous traction condition, i.e.
conservation of momentum, gives a discontinuous radial derivative of the
displacement
\begin{eqnarray}
\xi^\varphi_{\,\mathrm{core},r} = \left( 1+
\frac{\mu_\mathrm{S}}{\Phi^4 (b^r)^2}
\right)\xi^\varphi_{\,\mathrm{crust},r}\,. \label{eq_interface}
\end{eqnarray}
This condition is guaranteed by a particular reconstruction that relates the 
radial derivatives at both sides of the core-crust interface 
\citep{Gabler2012}. In the superfluid case, however, this particular 
reconstruction 
fails when the local Alfv\'en speed is larger in the core than the propagation 
speed in the crust. In this case, we abandon this particular reconstruction and 
use the usual reconstruction of our MHD scheme.
No additional conditions are necessary at the crust-core interface.

As mentioned in the previous section, superfluid neutron vortices, which exist 
due to the (slow) rotation, could be pinned to the crust, and, hence, 
introduce an additional coupling between the core matter and the crust. 
However, the amount of matter organised in the neutron vortices is negligible 
compared to the matter oscillating in the core or in the crust. Therefore, we account for 
our ignorance about the detailed vortex pinning and unpinning by the 
parameter $\varepsilon_\star X_c$.

\begin{figure}  
\includegraphics[width=.46\textwidth]{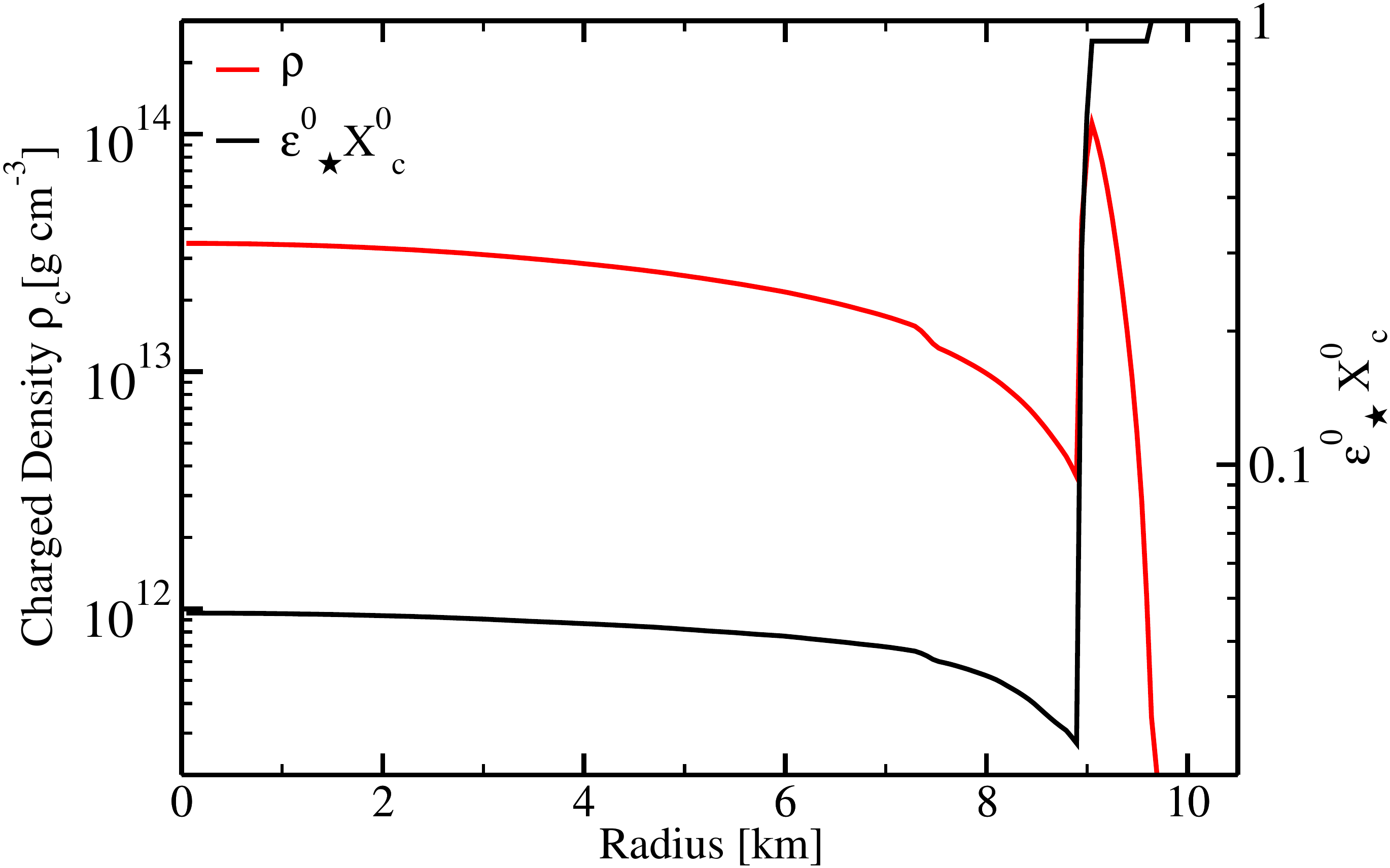}
\caption{Radial profiles of the density of the charged component and of the 
factor
$\varepsilon^0_\star X^0_c$ for our fiducial stellar model with a radius
$R=12.26\,$km ($r=10.08\,$km) and a mass $M=1.4\mathrm{M}_\odot$. Note that in 
this plot and throughout the paper we use the \textit{coordinate radius} $r=\Phi^2 R$. 
}
\label{fig_xc}
\end{figure}

We also do not treat pasta phases, which are probably present at the 
core-crust interface \citep{Chamel2008}, self-consistently. First, there is no
calculation of the shear modulus, fraction of charged particles (nuclei) and the
entrainment coefficients available for 
this state of matter. Second, we expect the main effect of the pasta
phases for the magneto-elastic oscillations to be a smearing out of the sharp 
transition of the core-crust transition \citep{Gearheart2011}. It is very 
likely that the shear  modulus does not vanish discontinuously as we are 
assuming. In particular, for pure shear, $n>0$ oscillations, this may have an 
effect, because their frequencies are defined by their travel time in the 
radial direction. However, for magneto-elastic QPOs the effects are expected to 
be minor: the shear modulus is anticipated to decrease inside the pasta phases 
and in this region the magnetic field will dominate, as it does in the absence 
of the pasta phases. 
One would also expect that the entrainment coefficients do not change from
strong entrainment inside the crust to weaker entrainment in the core
discontinuously, and similarly the charged particle fraction should not decrease
discontinuously at the core-crust interface. We thus decrease 
$\varepsilon_\star X_c$ continuously from the 
crust to the core
, see Fig. \ref{fig_xc}. 

\subsection{Magnetic field strength}\label{sec_bfield}

The definition of the magnetic field strength varies in the literature between 
different authors. \cite{Colaiuda2011} and \cite{vanHoven2012} used the polar 
magnetic field 
strength, \cite{Passamonti2013} took the average magnetic field in 
the stellar volume, while observational astronomers usually refer to the 
equivalent magnetic field of a rotating uniformly magnetised sphere 
\citep[see e.g.][]{Olausen2013}. In our previous 
work we have not been consistent in that respect either: in 
\cite{Gabler2011letter,Gabler2012} we used the magnetic field strength at the 
pole. In \cite{Gabler2013a} we also refer to the polar value and, to better 
compare with the spin-down estimate, we used in addition an 
equivalent magnetic field strength of a 
uniformly magnetised sphere that would cause the same spin-down as in our model. 
In \cite{Gabler2013b} we take the average surface magnetic field strength as 
reference.

Ultimately, we want to compare our magnetic field estimates directly to those of the 
spin-down estimates.  Therefore, similarly to \cite{Gabler2013a} we use here the 
magnetic field strength of a uniformly magnetised sphere that has the same 
magnetic dipole moment as the corresponding model, i.e. both our model and 
the sphere would have the same spin-down. However, in \cite{Gabler2013a}, we 
did not rescale the magnetic field strength to a sample model of a neutron
star with a radius $R=10\,$km, which we will do here. The equivalent magnetic
field strength is thus 
\begin{eqnarray}
\bar B = \frac{m}{(10\,\mathrm{km})^3} \left(\frac{10\,\mathrm{km}}{R}\right)^2\,,
\end{eqnarray}
where $m$ is the magnetic dipole moment of the equilibrium configuration and the
second factor corrects for the moment of inertia in the spin-down formula.
For our fiducial model ($R=12.26\,$km, $M=1.4\mathrm{M}_\odot$) the equivalent
magnetic field strength is roughly half of the field strength at the polar axis
$\bar B \sim 0.56 B_\mathrm{pole}$.

\section{Low frequency oscillations with realistic entrainment}\label{sec_low_f}

We study the superfluid magneto-elastic oscillations of our 
background magnetar model which is calculated with the {\tt RNS} code. It 
has a mass $M=1.4\mathrm{M}_\odot$ and a radius $R=12.26\,$km.
As described in the previous section, we take into account a realistic 
EoS (APR+DH) including the stratification resulting from a variable 
proton fraction in the core and the structure of the solid crust. We 
further consider the entrainment between protons and neutrons, which strictly 
speaking should vanish in models with non-superconducting protons. In this 
section we describe the features of the oscillations caused by the presence of 
superfluid neutrons and we study their behaviour at different magnetic field 
strengths. The resolution for the numerical simulations is $150\times80$ zones 
for 
$r\times\theta=[0\,\text{km},10\,\text{km}]\times[0,\pi]$.

\subsection{Classification of different magneto-elastic oscillations}
\begin{table*}
\begin{tabular}{c c c c c c c c}
Oscillation&specification&continuum (C)& Gabler et al. & 
Colaiuda 
et al. & 
Passamonti&vanHoven \&& this work\\
type&&or 
discrete (D)&Cerd\'a-Dur\'an et al.& Sotani et al.& et al.&Levin\\\hline
Alfv\'en/magneto-elastic  & upper & C & $U^{(\pm)}_n$ & 
$U^\mathrm{odd/even}_n$ & $U^{(\pm)}_n$ &turning point& $U_n^\mathrm{surf}$ 
, $U_n^\mathrm{ 
core } $\\
Alfv\'en/magneto-elastic & edge & C & $E^{(\pm)}_n$ & 
$L^\mathrm{odd/even}_n$ & $L^{(\pm)}_n$&edge&$E_n$\\
Alfv\'en & lower  & C & $L^{(\pm)}_n$ & $C_n$ & $C_n$ &turning 
point& $L_n$\\
shear mode & & D & crustal & crustal & $^lt_n$ &crustal& $^lt_n$\\
discrete magneto-elastic& & D & - & discrete Alfv\'en & $^lt_n^\ast$ 
&elasto-magnetic
& 
$^lU_n$\\
\hline\hline
\end{tabular}
\caption{Comparison of the notation used by different groups for the 
classification of oscillations. 
Here (last column), we continue employing the notation of Upper (U), Edge (E), 
and Lower (L) oscillations
as in our previous work \citep{Cerda2008,Gabler2012,Gabler2013a,Gabler2013b}. 
However, we will omit the symmetry label $(\pm)$ and introduce 
some additional specifications. (See main text for 
details.)}\label{tab_QPOs_notation}
\end{table*}

\begin{figure}  
\includegraphics[width=.44\textwidth]{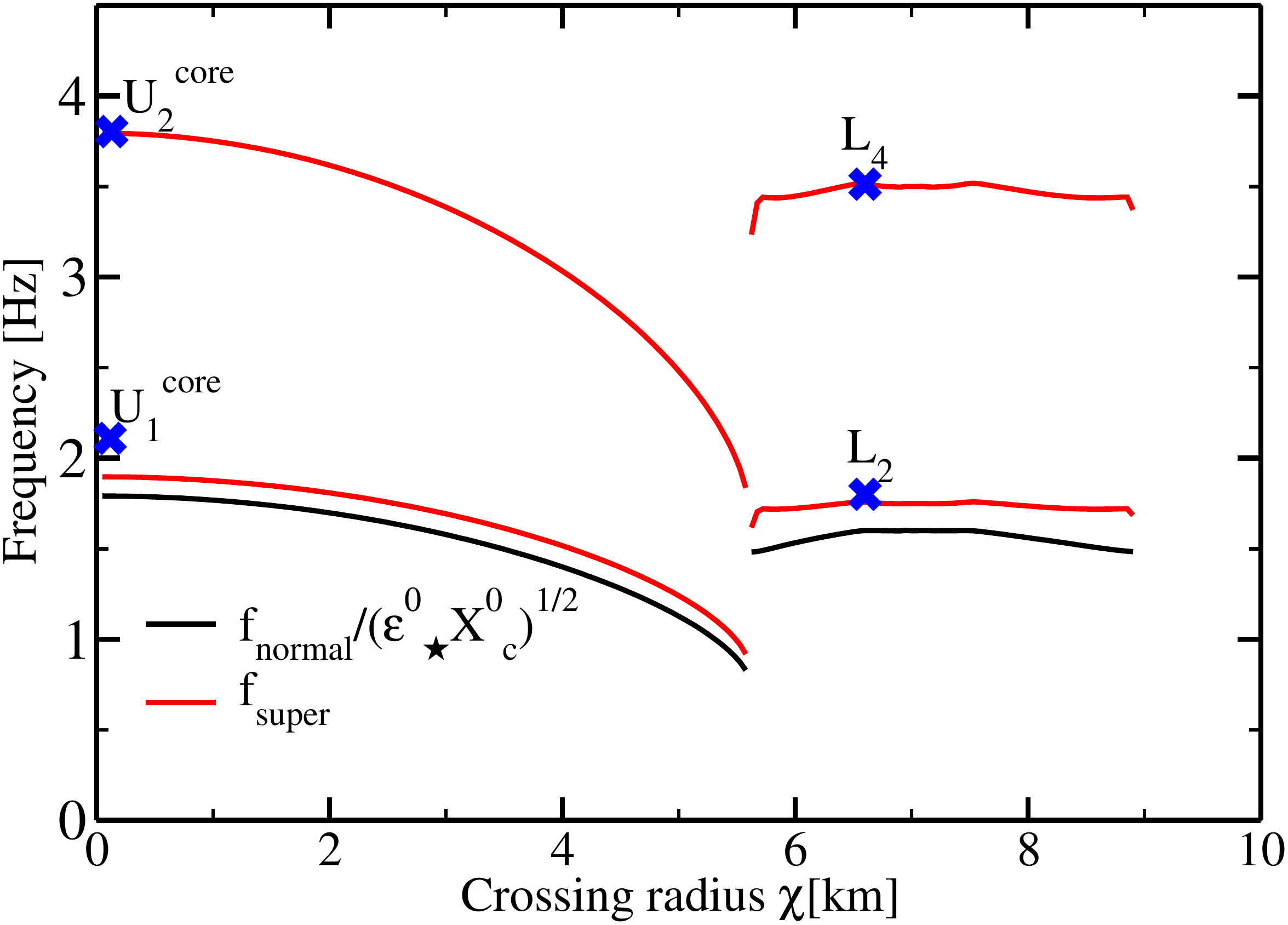}
\caption{Semi-analytic, low-frequency spectrum of several 
continuum magneto-elastic oscillations for our magnetar model at $\bar 
B=5\times10^{13}\,$G. The red lines include superfluid effects, while the black 
lines and correspond to the rescaled spectrum without 
superfluidity. The corresponding oscillation frequencies obtained in the 
simulations including superfluidity are indicated by blue crosses
(compare also with 
Fig.\,\ref{fig_QPO_5_13}).}
\label{fig_spectrum}
\end{figure}

As we discuss below, new structures in the QPO families appear in our model as a 
consequence of the inclusion of superfluidity. Therefore, to accommodate for the 
new oscillations we 
need to improve the notation employed for the classification of the oscillations, 
both compared to
other work and to {\it our} previous 
work \citep{Gabler2012,Gabler2013a,Gabler2013b}. In these references we 
classified the oscillations according to their location within the Alfv\'en 
continuum (see also 
Table\,\ref{tab_QPOs_notation}):
\begin{itemize}
 \item[U:]Upper QPOs that are related to the upper turning point of the 
spectrum, close to the polar axis.
\item[E:]Edge QPOs that are located at the edge of the spectrum close to the 
region of closed field lines near the equatorial plane.
\item[L:]Lower QPOs that appear in the closed field line region.
\end{itemize}
As an example, we plot in Fig.\,\ref{fig_spectrum} a spectrum for our model with
$\bar B=5\times10^{13}\,$G as a function of the radius $\chi$ at which the 
corresponding field line crosses the equatorial plane. The spectrum is obtained 
with our semi-analytic 
model \citep{Cerda2008,Gabler2012} which is based on the integration of the 
Alfv\'en speed along a given magnetic field line and by assuming perfect 
reflection of the oscillations at the core-crust interface. In this example, the 
Upper QPOs are located at $\chi=0$ km, the 
Edge QPOs at $\chi\sim5.5\,$km and the Lower QPOs at $\chi\sim6.5\,$km 
(confined to the region of magnetic field lines that close inside the core, 
$6\,{\rm km}<\chi <9\,$km). The larger deviation for $U_1^\mathrm{core}$ 
is expected, because of the limited accuracy of the semi-analytic model for 
long wavelengths \citep{Cerda2008}.

\begin{figure*} 
\includegraphics[width=\textwidth]{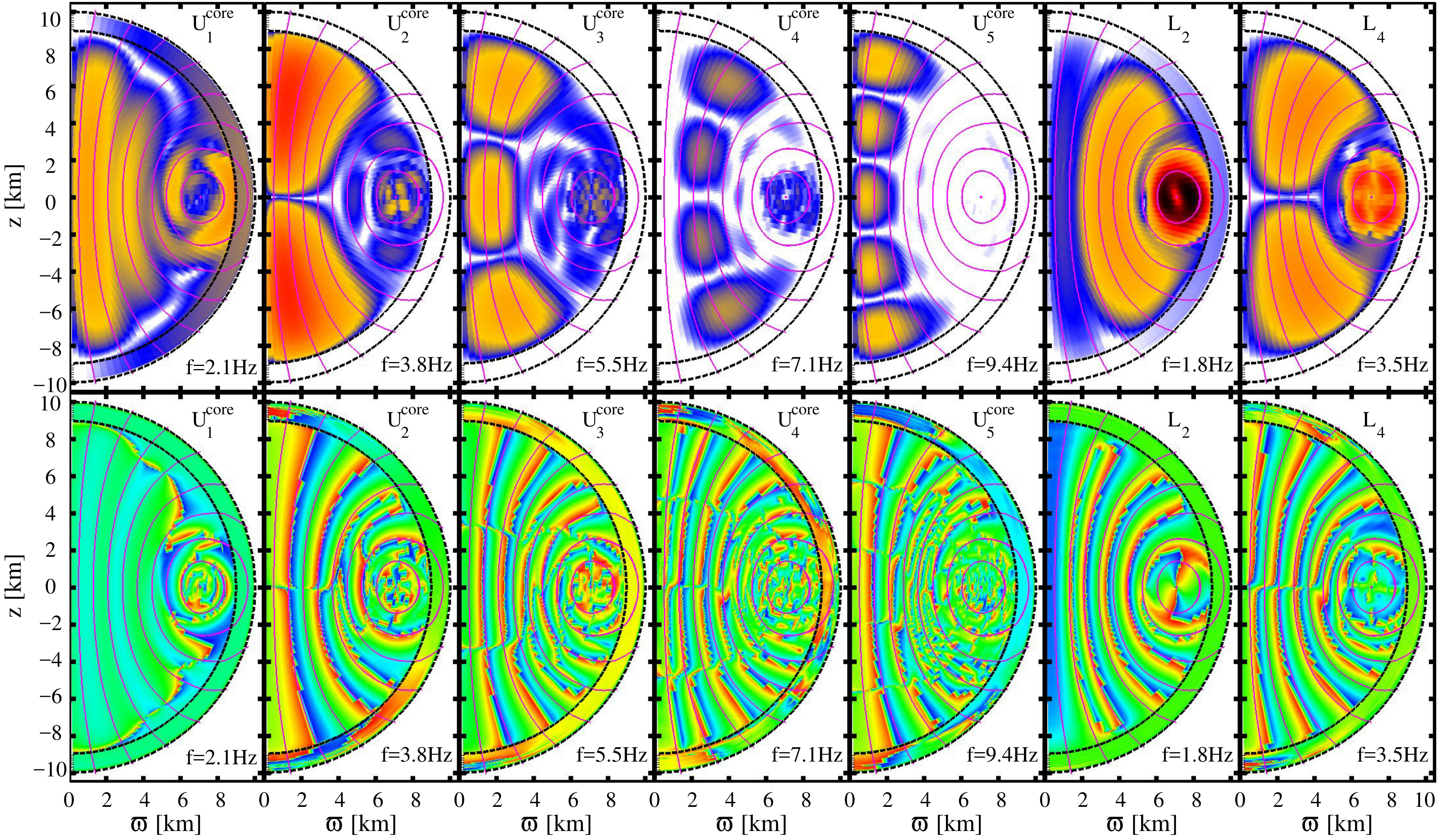}
\caption{Amplitude and phase distributions (resulting from the Fourier transform 
of the velocity evolution) for the strongest QPOs 
with frequencies up to $\sim10\,$Hz, for $\bar B=5\times10^{13}\,$G .
{\it Top row}: Amplitude distribution. The colour scale ranges 
from white-blue
(minimum) to red-black (maximum). {\it Bottom row}: Phase distribution. 
The colour scale ranges from blue ($\theta=-\pi/2$) to 
orange-red ($\theta=\pi/2$). The crust is indicated by the region between the 
dashed black 
lines and magnetic field lines are plotted in magenta. Notice 
that the plots for the ${\rm L}_2$ and ${\rm L}_4$ oscillations contain residual 
amplitudes outside the region of closed magnetic field lines, due to other, 
nearby QPOs. Also, sudden jumps from red to blue colour in the phase 
distribution 
are no discontinuities, but are caused by the phase crossing multiples of $\pi$ 
(the phase changes continuously throughout the star by several such multiples).}
\label{fig_QPO_5_13}
\end{figure*}

To allow for a unique identification of the members of each of the QPO classes 
U, E, and L we had previously introduced a superscript + (-) 
for QPOs that are symmetric 
(antisymmetric) with respect to the equatorial plane, and a subscript 
$n\in\{0,1,2,\dots\},$ which indicated the fundamental $(n=0)$ oscillation and 
the respective overtones $(n\geq1)$. For example, the members of the class of 
symmetric upper QPOs were denoted as $U^+_0$, $U^+_1$, 
$U^+_2$, $\dots$\,.
To better discriminate the different QPOs, we introduce here additional 
specifications, such as $U_n^\mathrm{core}$ and  $U_n^\mathrm{surf}$ to indicate
oscillations $U$ that are confined to the core or reach the surface, 
respectively. Moreover, in this 
work $n$ gives the {\it number of maxima} along the field lines, but not the 
overtone in a particular family. This notation is unique for oscillations 
that are symmetric or antisymmetric with respect to the equatorial plane, and therefore 
we will omit the explicit indication of symmetry $(\pm)$ from now 
on. An even $n$ directly 
implies antisymmetry, while an odd $n$ implies symmetry.
We also introduce discrete magneto-elastic oscillations $^lU_n$ in analogy to 
pure crustal shear modes $^lt_n$. For the latter, $l$ is the 
spherical harmonic index in the $\theta$-direction and $n$ is the number of 
maxima in the radial direction. Correspondingly, for the discrete 
magneto-elastic oscillations, we use $l$ as the number of maxima inside the 
crust \textit{across} magnetic field 
lines and $n$ as the number of maxima throughout the star \textit{along} the 
magnetic field lines in the region where the oscillation dominates.

\subsection{General description of the oscillations}

In models without superfluidity, the observed low-frequency QPOs were
explained as magneto-elastic oscillations \citep{Gabler2011letter,Gabler2012, 
Gabler2013a}. For very weak magnetic fields of $\bar B\lesssim 
10^{13}\,$G, torsional shear oscillations were obtained, which could explain 
some of the  observed QPO frequencies. However, these oscillations were 
efficiently damped 
into the core for magnetic fields strengths in the range $10^{13}\lesssim 
\bar B\lesssim10^{15}\,$G. In this case, Alfv\'en oscillations in the core 
were reflected at the core-crust interface and the oscillations were confined 
to the core. At $\bar B\gtrsim10^{15}\,$G, the oscillations could penetrate 
into
the crust and reach the surface. For both magnetic field regimes the 
oscillations had similar 
structures with one up to several maxima along the magnetic field lines and no 
particular structure perpendicular to the field lines \citep[see e.g. figure 10 
in][]{Gabler2012}.

Including superfluid effects, we find that the behaviour of the oscillations at 
low magnetic field strengths ($\bar B<10^{14}\,$G) is very similar to 
non-superfluid models. The structure of the dominant oscillations is shown in  
Fig.~\ref{fig_QPO_5_13} for $\bar B=5\times10^{13}\,$G, which also shows the magnetic
field lines in magenta colour.
The main difference superfluidity brings is that the oscillation frequencies are scaled upwards
roughly by the average of $\varepsilon^0_\star X^0_c $ along 
the corresponding magnetic field line. In the top row of 
Fig.\,\ref{fig_QPO_5_13} we plot the magnitude of the complex
Fourier amplitudes of the velocity of the perturbation for each point of our 
numerical grid at the frequencies of 
the first five Upper oscillations, 
$U_1^\mathrm{core}$ to $U_5^\mathrm{core}$, which are confined to the core, and 
the two Lower 
oscillations $L_2$ and $L_4$. The colour scale 
spans from white-blue (minimum) to red-black (maximum).
In the bottom row of Fig.\,\ref{fig_QPO_5_13} we show the corresponding phase 
obtained from the Fourier analysis.
Here, the colour scale spans from $-\pi/2$ (blue) to $+\pi/2$ (orange-red). 
Jumps from red to blue are no discontinuities, but are caused by the phase 
crossing multiples of
$\pi$. For example, in the region of open field lines, the phase changes continuously throughout the star by several such
multiples, i.e.~$\phi=\phi_0+N\pi$, where $N$=1, 2, .... and $\phi_0$ is the 
phase at along the axis.  More specifically, for 
$U_2^\mathrm{core}$ (second column in Fig.\,\ref{fig_QPO_5_13})  the phase 
changes from green ($\phi=\phi_0$) to  red/blue ($\phi=\phi_0+\pi/2$), again to green 
($\phi=\phi_0+\pi$) and  to red ($\phi=\phi_0+3\pi/2)$, etc. In the region of 
open field lines a  total phase change of roughly $\Delta\phi\sim 7\pi/2$ can be 
observed for this oscillation. Similarly, all Upper oscillations 
$U_n^\mathrm{core}$ show a continuous change of phase in the region of open 
field lines, which underlines their characterization as continuum oscillations, 
rather than being discrete
modes. Notice that in the region of open field lines confined to the 
core, the 
phase does not change along individual magnetic field lines. 
In contrast, the $L_2$ and $L_4$ oscillations  in the region 
of closed field lines,  show a varying 
phase along  the field lines.(The resolution in 
these simulations was not sufficient to clearly show the phase changes for some 
oscillations).

All of the properties of the 
oscillations described in the previous paragraph are also present in  non-superfluid models.
The frequencies obtained for the $U_n^\mathrm{core}$ and $L_n$ oscillations 
agree very well with our semi-analytic model, see Fig.\,\ref{fig_spectrum}.
When including superfluid effects, major differences arise for magnetic fields 
stronger than $10^{14}\,$G. Firstly, the lowest magnetic field strength at 
which the oscillations can reach the surface is reduced significantly to $\bar 
B\gtrsim {\rm few}\times 10^{14}\,$G compared to $\bar B\gtrsim10^{15}\,$G 
in the normal fluid case
\citep{Gabler2012}. Similarly, \cite{vanHoven2011} found a strong 
damping of crustal motion for magnetic field strengths $\bar B < 10^{15}\,$G, 
if the core neutrons are normal. Secondly, the oscillations now have 
structures also 
perpendicular to the magnetic field lines.  

\begin{figure}  
\includegraphics[width=.46\textwidth]{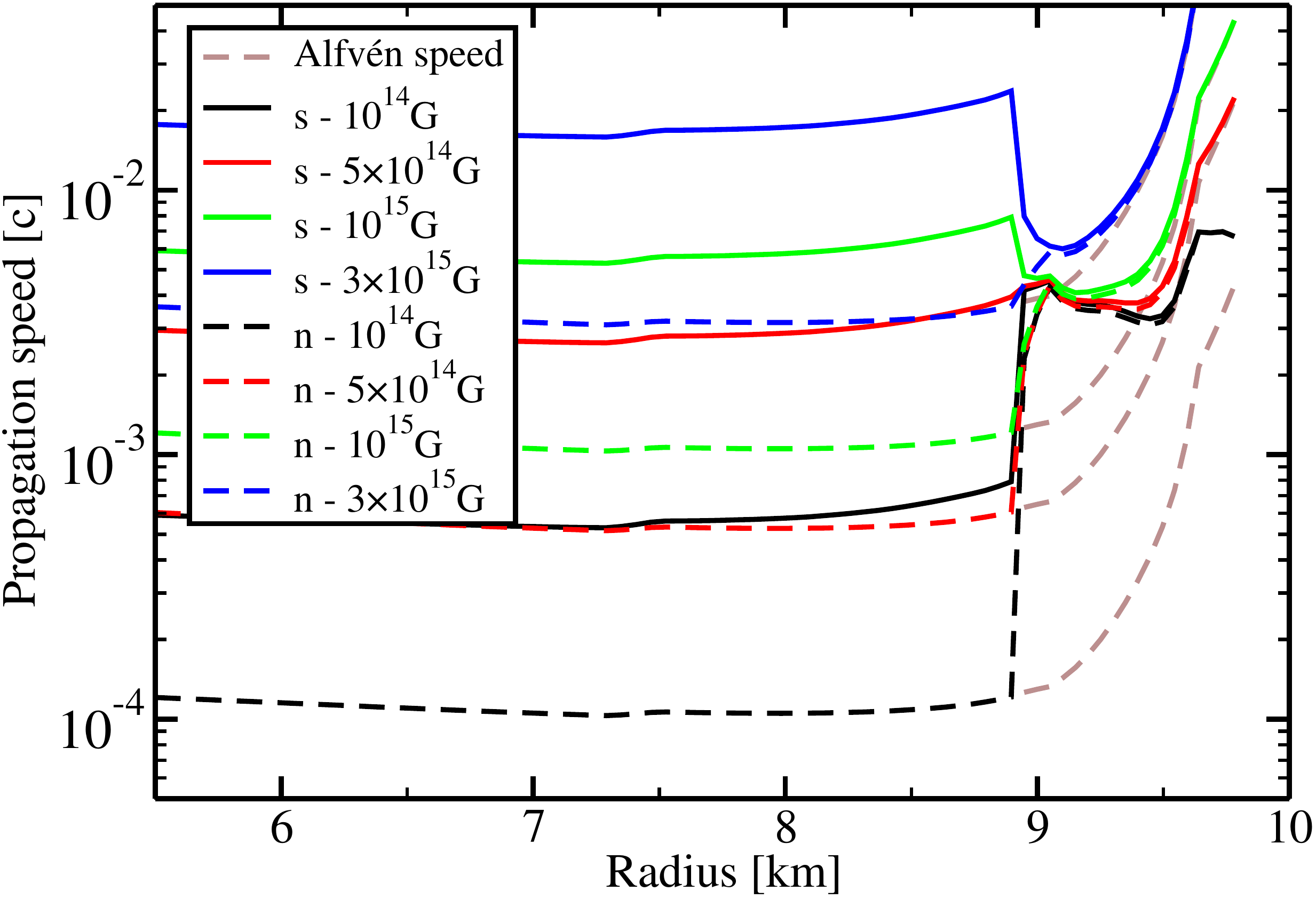}
\caption{Radial profiles of the propagation speeds of  
magneto-elastic oscillations for different
magnetic field strengths, close to the polar axis. Solid lines correspond to superfluid 
neutrons and dashed lines to normal matter. The crust starts at $\sim 9\,$km, 
and 
the brown dashed lines indicates the Alfv\'en velocity in the crust.}
\label{fig_speeds}
\end{figure}

The penetration of oscillations into the crust region depends on the 
impedance mismatch between crust and core. To some extend this is related to 
the  relative  amplitude of the Alfv\'en velocity compared to that of the 
shear velocity in the  crust \citep[see ][for a more detailed 
discussion]{Gabler2012, Link2014}.
In Fig.~\ref{fig_speeds} we plot the different 
propagation velocities 
near the polar axis for different magnetic field strengths and for simulations 
with (solid lines) and without (dashed lines) superfluid neutrons. For very low 
magnetic fields, $\bar B\lesssim10^{14}\,$G, the Alfv\'en velocity in the core 
is much smaller than the shear 
velocity at the base of the crust and the impedance mismatch is large. In 
this case, most incoming waves are reflected at the core-crust interface. The 
interface acts like a solid 
wall, where the waves get reflected back into the interior. The 
reflected oscillations stay confined to the core and have a node at the 
core-crust interface (see top row of Fig.~\ref{fig_QPO_5_13}). With increasing 
magnetic field strength the difference in propagation velocity 
and impedance decreases and not all of the waves are reflected. Therefore, 
the oscillations start to have non-negligible amplitudes in the crust. 

Fig.~\ref{fig_speeds} also shows that in the superfluid case the Alfv\'en velocity is 
higher than in the normal fluid case for a given magnetic field strength. Hence, the 
propagation velocities at both sides of the crust-core interface differ 
less and also the impedance mismatch is smaller. These smaller 
differences allow the oscillations to reach the surface at lower magnetic 
field strengths if 
neutrons are superfluid. For example, at $5\times10^{14}\,$G the Alfv\'en 
velocity in 
the core equals the propagation speed in the crust (solid red line in Fig.\,\ref{fig_speeds}), 
and there should be weak reflection for the considered field line 
compared to weaker magnetic fields. 

\begin{figure*} 
\includegraphics[width=\textwidth]{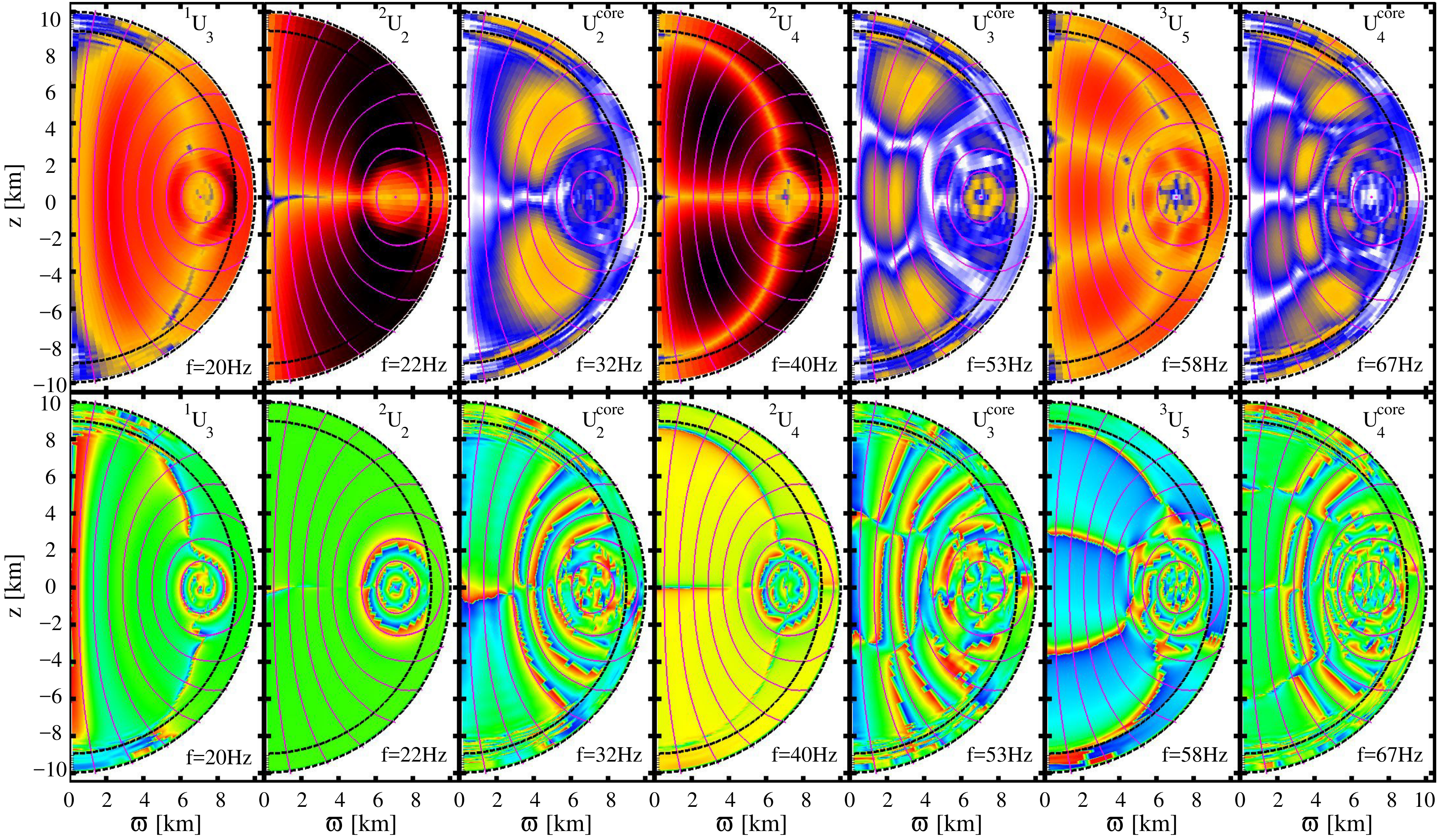}
\\[1cm]
\includegraphics[width=\textwidth]{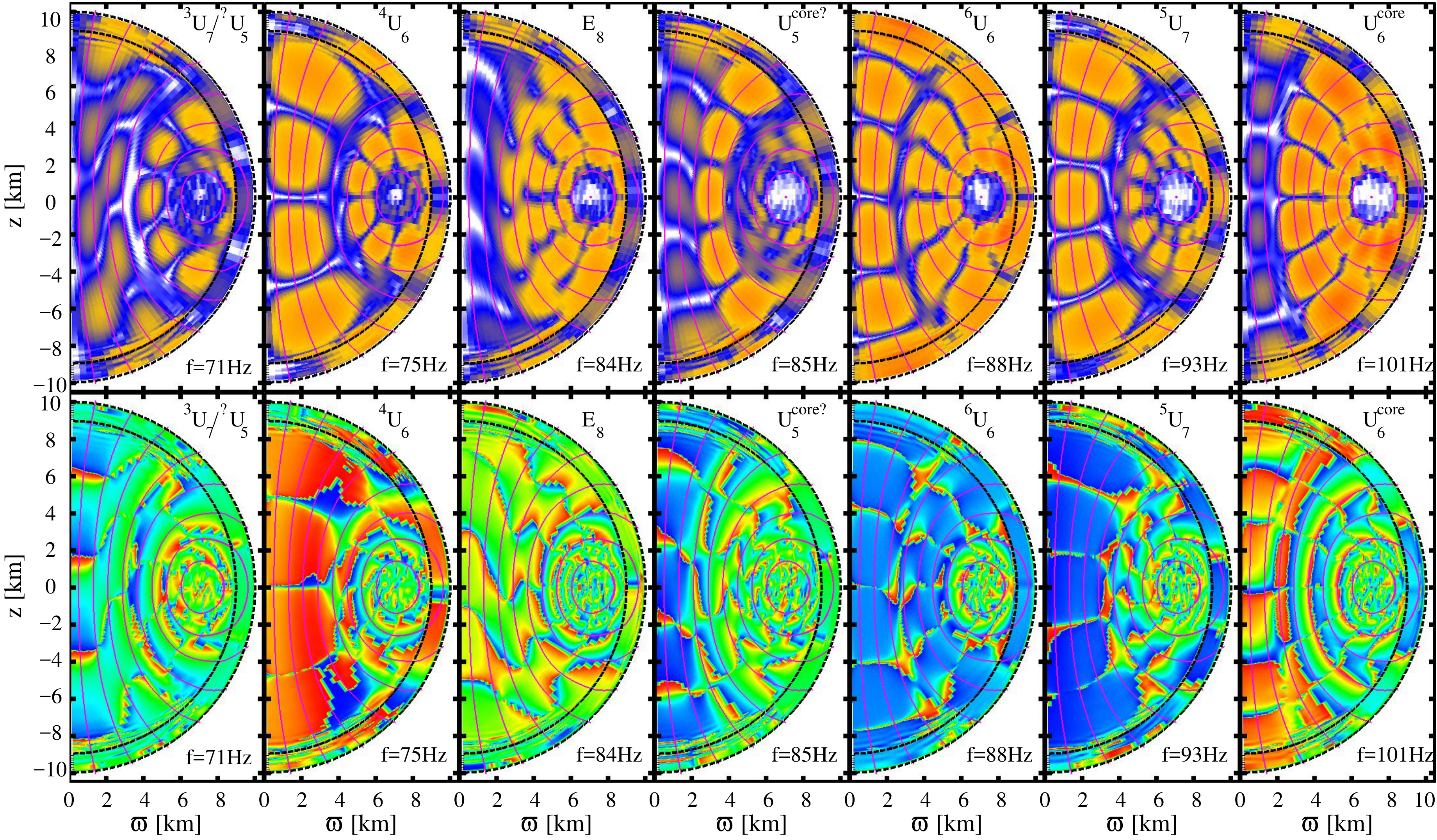}
\caption{
Discrete Fourier transform of the velocity of the strongest QPOs 
at $\bar B=5\times10^{14}\,$G with frequencies up to $\sim100\,$Hz.
{\it First and third row}: Fourier amplitude. {\it Second and fourth row}: 
Corresponding phases.
Colour scales as before in Fig.\,\ref{fig_QPO_5_13}.}
\label{fig_QPO_5_14}
\end{figure*}

Our simulations exhibit the expected behaviour, as can be seen in Figs.~\ref{fig_QPO_5_13} 
and \ref{fig_QPO_5_14}. At $\bar{B}=5\times10^{13}\,$G (Fig.\,\ref{fig_QPO_5_13}), all 
oscillations are confined to the core, while at $\bar{B}=5\times10^{14}\,$G 
(Fig.\,\ref{fig_QPO_5_14})  oscillations like $^2U_2$, $^2U_4$, and 
$^3U_5$, display a strong magneto-elastic character right up to the 
surface. At this magnetic 
field strength, other oscillations (such as  $U_2^\mathrm{core}$, $U_3^\mathrm{core}$, 
$U_4^\mathrm{core}$) show only a partial excitation in the 
crust with an amplitude distribution that is similar to that 
of shear oscillations. 
This is due to partial reflection of magnetic field lines which are far from 
the pole, since i) there is still a  mismatch of the impedances at the 
core-crust interface , and ii) the angle changes under which 
the magnetic field enters into the crust.

\begin{figure*} 
\includegraphics[width=\textwidth]{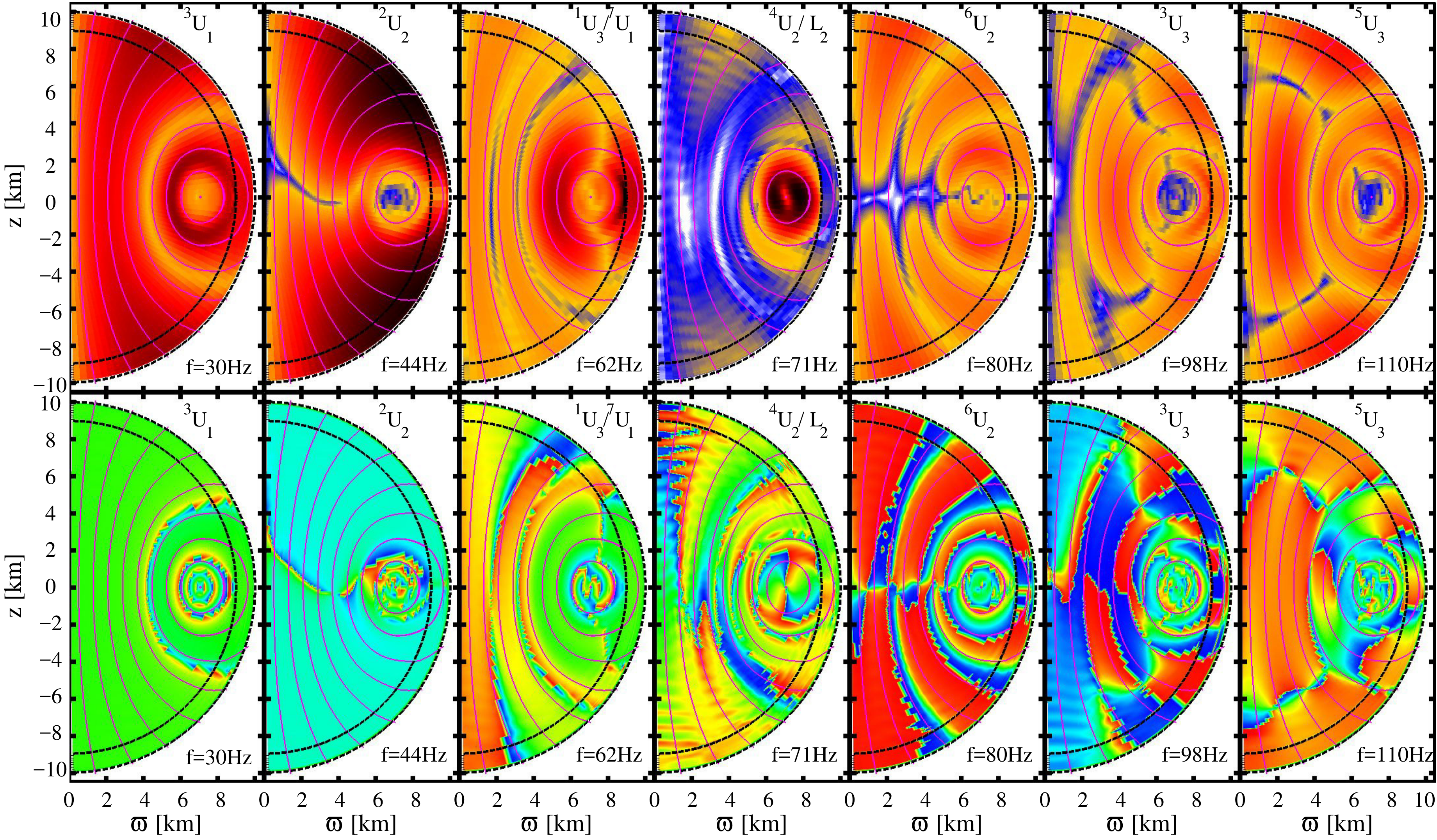}
\\[1cm]
\includegraphics[width=0.72\textwidth]{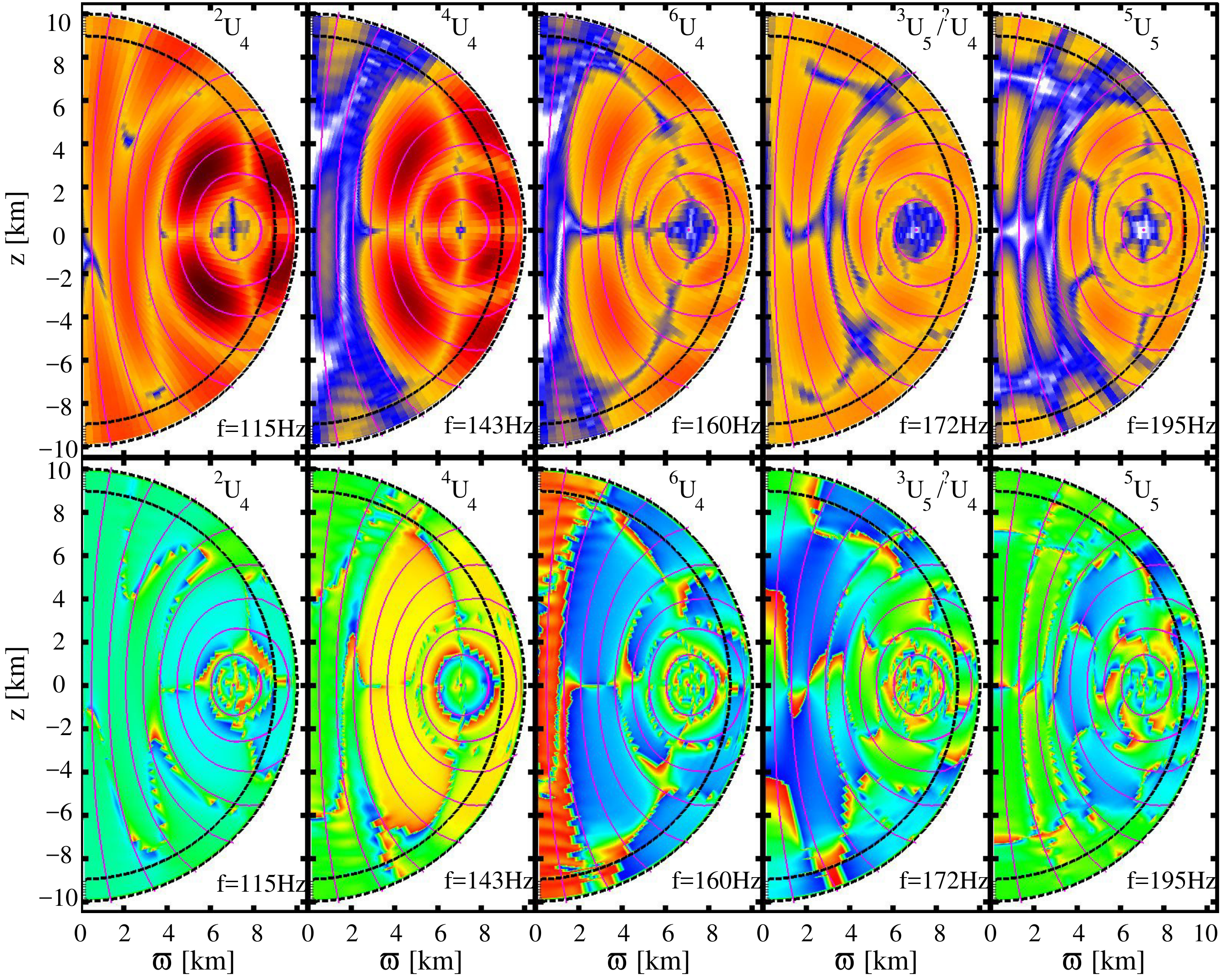}
\caption{Discrete Fourier transform of the velocity of the strongest QPOs 
at $\bar B=2\times10^{15}\,$G with frequencies up to $\sim200\,$Hz.
{\it First and third row}: Fourier amplitude. {\it Second and fourth row}: 
Corresponding phases.
Colour scales as before in Figs.\,\ref{fig_QPO_5_13} and \ref{fig_QPO_5_14}.
}
\label{fig_QPO_2_15}
\end{figure*}

In addition to the different behaviour regarding the reflection at the core-crust 
interface, the structure and the properties of the oscillations also change 
when including superfluid effects. To discuss this, we will use 
Figs.\,\ref{fig_QPO_5_14} and \ref{fig_QPO_2_15} which show the results of the 
Fourier transform at the frequencies of the dominant oscillations for 
$\bar{B}=5\times10^{14}\,$G and $\bar{B}=2\times10^{15}\,$G, respectively. 
 The simulation time was $1.350\,$s and $0.338\,$s, respectively which 
corresponds to about 80 oscillation periods of the lowest frequency 
oscillation.
As in Fig.\,\ref{fig_QPO_5_13}, the top row in both panels show the Fourier 
amplitude 
and the bottom row the corresponding phase, and we employ the same colour scale. 
At $5\times10^{14}\,$G (Fig.\,\ref{fig_QPO_5_14}), there are two distinct
 groups of oscillations, one that consists of oscillations showing an Alfv\'en
character confined to the core and a very weak shear-oscillation 
character in the crust (such as $U_2^\mathrm{core}$ to $U_6^\mathrm{core}$), 
and the other displaying a strong magneto-elastic 
character up to the surface ($^1U_3, ^2U_2, 
^2U_4, ^3U_5, ^6U_6)$.

In addition, the two groups have different phase properties. The 
phase of oscillations  belonging to the first group, is continuous in the 
region of their largest amplitudes. For the second 
group with the strong magneto-elastic character up to the surface, the phase is 
very nearly constant $(^2U_2, 
^2U_4,  ^6U_6)$ or only shows a small variation $(^1U_3, 
^3U_5)$. It is important to note that at nodal lines, the 
computation of the phase has large numerical errors and the constancy of the 
phase should only be judged by its value in regions for from nodal lines. 

However, there are some oscillations like $^3U_7, ^4U_6,$ and $ ^5U_7$, that 
seem not to match clearly into one or the other group. They have intermediate 
amplitudes inside the crust, and their phase varies in different parts of the 
neutron star. For $^3U_7$, which has its maximal amplitudes between the 6th and 
7th plotted field line (magenta lines) the phase in this area and the part of 
the crust near by the equatorial plane is almost constant. Only close to the 
polar axis, where an overlap with a $^?U_5$ oscillation appears, the phase 
varies modestly. Oscillations $^4U_6$ and $^5U_7$ show a nearly constant phase 
in their dominant regions close to the polar axis and inside the crust. Notice 
that for these two oscillations there is an overlap with edge modes excited at 
the last open field line, which have a different phase and a different number of 
nodal lines than the dominant part of the oscillation. In all these three cases 
the phase of the oscillation actually helps to identify the dominating 
oscillation by indicating which parts of the neutron star oscillate together and 
have the same phase. Problems for a clear identification of the oscillations are 
small amplitudes compared to the dominant oscillations and limited resolution 
for oscillations with increasing $l$ and $n$. Additionally, a high $n$ and low 
$l$ oscillation may have a very similar frequency compared to a low $n$ but 
high $l$. Then both oscillations may appear in the same plot of a Fourier 
amplitude complicating a clear characterization (see e.g. $^3U_7$ and $^?U_5$ 
in Fig.\,\ref{fig_QPO_5_14}).

For stronger magnetic fields, $\bar B\sim2\times10^{15}\,$G, the picture changes drastically and we only find 
coherent oscillations that reach the up to the surface. In the 
corresponding Fig.\,\ref{fig_QPO_2_15}, we also see that the oscillations show 
structure perpendicular to the field lines. In contrast, for non-superfluid 
models \citep{Gabler2012} the oscillations have a simpler structure with only 
one maximum perpendicular 
to the field lines. A similar situation appears in the superfluid case 
when considering a weaker 
magnetic field, $\bar B\sim5\times10^{13}\,G$; compare Fig.\,\ref{fig_QPO_5_13} in this
work with Fig.~10 in \cite{Gabler2012}. For the even stronger magnetic fields analysed, $\bar 
B>5\times10^{15}\,G$, the oscillations become again dominated by the magnetic 
continuum, they have continuous phases and a simple structure with only one 
maximum perpendicular to the magnetic field lines. In fact, they look very 
similar to the oscillations in Fig.\,\ref{fig_QPO_5_13} that are located near 
the polar axis. However, for stronger magnetic fields they exhibit a maximum at 
the surface analogous to normal fluid models.

\subsection{ Coherent oscillations}

We turn now to discuss coherent oscillations with constant phase.
They only exist 
above a critical magnetic field strength, reach the magnetar 
surface and have much higher amplitudes 
than those of the continuum oscillations of the core. Inside the crust, 
the coherent oscillations look somewhat similar to pure crustal modes. For 
example,  
$^2U_2$ and $^2U_4$ resemble $^2t_0$, or $^3U_5$ resembles 
$^3t_0$ (cf.~Fig.\,\ref{fig_QPO_5_14}). 
Similar oscillations have been reported by 
\cite{vanHoven2011,vanHoven2012}, who found shear modes that seemed to be 
shifted into the gap of the continuum of the core if the crust was fixed.
Discrete (Alfv\'en) modes were found in the simulations of~\cite{Colaiuda2011} 
in models without superfluid effects. These modes were also (partially) 
interpreted as crustal modes that exist in the gap of the continuum of the 
core.

The coherent oscillations with constant phase found here may be related 
to these `gap modes',
although we interpret them in a different way. They should not be 
mistaken for shifted pure crustal shear modes for the following reasons: 
(i) Their frequencies do not match the shear mode frequencies of the crust, which
are given in Table\,\ref{tab_shear_modes}. For example at 
$\bar{B}=5\times10^{14}\,$G the frequencies of the coherent oscillations are 
$f_{^2U_2}=22\,$Hz, $f_{^2U_4}=40\,$Hz and $f_{^2t_0}=26.5\,$Hz, or 
$f_{^3U_5}=58\,$Hz and $f_{^3t_0}=41.9\,$Hz.
(ii) Different coherentoscillations have similar structure within 
the crust, e.g.~$^2U_2$ and $^2U_4$ at $\bar B=5\times10^{14}\,$G look very similar 
to $^2t_0$ (inside the crust).
(iii) When increasing the magnetic field strength further, the structure of the oscillations inside 
the crust can deviate significantly from the structure of crustal modes. A good 
example is provided by the oscillations $^2U_4$,$^4U_4$ and $^6U_4$ at $\bar 
B=2\times10^{15}\,$G shown in Fig.\,\ref{fig_QPO_2_15}.

\begin{table}
\begin{tabular}{c c c c c c c }
$n$&0&0&0&0&0&1\\
$l$&2&3&4&5&6&2\\\hline
$^lf_n$&26.5&41.9&56.2&70.1&93.8&782.6\\
\hline\hline
\end{tabular}
\caption{Frequencies (in Hz) of the crustal shear modes of our fiducial model.}
\label{tab_shear_modes}
\end{table}

We interpret our simulations as follows. For weak
magnetic fields, $\bar B\lesssim\mathrm{few}\,10^{14}\,$G, the oscillations 
are confined to the core, crustal modes are damped efficiently, and the dominant
oscillations are Alfv\'en oscillations of the continuum. For stronger fields, a 
few $10^{14}\,\mathrm{G}\lesssim\bar B \lesssim 
\mathrm{several}\,10^{15}\,$G, the difference of the corresponding 
propagation 
velocities at the  core and the crust is small (Fig.\,\ref{fig_speeds}). Thus, 
the coupling of the crust to the Alfv\'en 
oscillations of the core is very efficient. Additionally, for such values of $\bar{B}$, the oscillations 
can enter into the crust where they are still dominated by the shear modulus. This 
leads to a strong interaction of oscillations of different field lines. This 
coupling is weaker in normal fluid models because, as mentioned before, the 
oscillations are reflected 
at the core-crust interface for significantly stronger magnetic fields than for superfluid models. 
Therefore, once they enter into the crust the oscillations are already of magneto-elastic type 
or dominated by the magnetic field inside the crust too. A strong coupling of 
continuum oscillations can lead to the formation of global modes, as 
\cite{Levin2007} and \cite{Cerda2008} showed for unphysically large 
dissipation and coupling through strong artificial viscosity, respectively. 
Normal modes are also found for highly tangled magnetic fields 
\citep{Link2015,Sotani2015}. Since we 
cannot show that the newfound 
oscillations show all properties of normal modes (like being orthogonal to each 
other and spanning a complete set), we call them {\it constant phase 
oscillations} (instead of discrete normal modes) to distinguish 
them from continuum oscillations. The new 
oscillations share properties of both, the Alfv\'en continuum oscillations 
(structures along the magnetic field lines extending into the 
core and frequency roughly scaling with $\bar B$, as we show below) {\it and} 
shear modes (structures orthogonal to field lines with constant phase). We thus 
conclude that the coherent oscillations are global magneto-elastic 
oscillations that are distinct from the purely shear modes of the crust. 

One particular characteristic of the coherent oscillations is their 
splitting in the  $\theta$-direction. The strong coupling caused by the 
crust 
allows for additional oscillation patterns of different angular 
dependencies (Figs.~\ref{fig_QPO_5_14} or \ref{fig_QPO_2_15}). This effect 
is reduced for very strong magnetic fields, i.e.~$\bar B$
larger than several $10^{15}\,$G, when the 
magnetic field dominates 
over the shear modulus inside the crust, and the spectrum is formed by 
continuum 
oscillations. Without superfluidity, on the other hand, the oscillations  
are reflected at the core-crust interface $\bar B \lesssim 10^{15}\,$G, a value 
that is actually comparable to the shear modulus inside the crust. Therefore, 
for normal fluid cores, the range of magnetic field strengths in which 
 coherent magneto-elastic oscillations can be observed is markedly 
narrowed down to about 
$10^{15}\,\mathrm{G}\lesssim \bar B \lesssim \mathrm{few}\times 
10^{15}\,\mathrm{G}$. We note that in our previous work with normal fluid models 
we did not fine tune our simulations to span such a narrow range of values 
of $\bar{B}$, which explains the fact that the coherent oscillations went 
unnoticed in our examination.

It should be emphasized that the discovery of these new oscillations generates 
a difficult new conundrum. Namely, explaining how these constant phase 
oscillations could account for the observed QPOs in the giant flares of two 
magnetars estimated to have significantly different magnetic field 
strengths, $\bar B\sim2.0\times10^{15}\,$G for SGR 1806-20 and $\bar B \sim 
7\times 10^{14}\,$G 
for SGR 1900+14, remains a challenging open issue.

\subsection{Scaling of the oscillation frequencies with $\bar B$} 
\label{sec_scaling_B}

Figs.\,\ref{fig_QPO_5_14} and \ref{fig_QPO_2_15} show 
that the oscillation frequencies depend on the magnetic field 
strength. In this Section, we investigate this dependence more systematically. 
To achieve this goal, we perform a new set of simulations with 
$\bar B=\{0.5,1,2.5,5,6,7,8,9,10,15,20,30\}\times10^{14}\,$G using the same initial 
data, namely a perturbation with a mixed $l=2$ and $l=3$ angular dependence and a radial 
structure having zero amplitude both in the centre and at the core-crust 
interface, 
 and reaching a maximum inside the core and at the surface of the magnetar.
This initial perturbation should excite many different oscillations 
that represent the spectrum of the magnetar model.

\begin{figure}
\includegraphics[width=.46\textwidth]{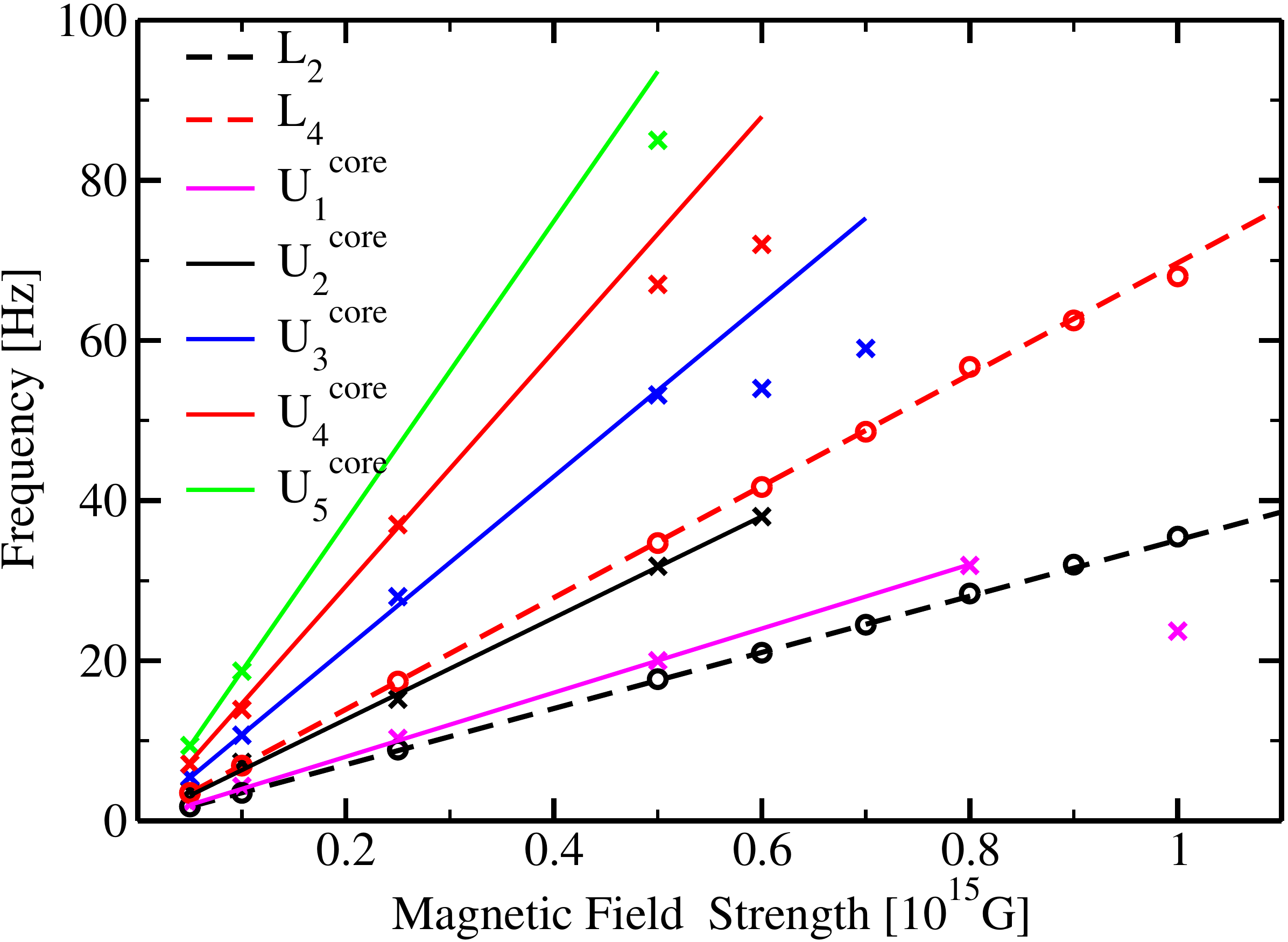}
\caption{Frequencies of the continuum oscillations having the lowest 
frequencies at as a function of magnetic field strength. Symbols give
the frequencies obtained from 
the simulations, and the lines are linear fits (see Table\,\ref{tab_fit_B1}).}
\label{fig_freqs1}
\end{figure}

\begin{table}
\begin{tabular}{c c c c c c c c}
  & 
$L_2$&$L_4$&$U_1^\mathrm{core}$&$U_2^\mathrm{core}$&$U_3^\mathrm{core}
$&$U_4^\mathrm{core}$&$U_5^\mathrm{core}$\\ \hline
$b$ [Hz]&3.5&7.0 &4.0& 6.3&10.8&14.7&18.7\\
\hline\hline
\end{tabular}
\caption{Linear fits of the frequency $f\,[\mathrm{Hz}] = a+ b \times \bar 
B[10^{14}\mathrm{G}]$ of the continuum oscillations of our superfluid model 
with realistic entrainment. For all Lower $L_n$ and Upper $U_n^\mathrm{core}$ 
oscillations that are confined to the core $a=0.0$.}
\label{tab_fit_B1} 
\end{table}

We first study the continuum oscillations. Their frequencies are shown as a 
function of the magnetic field strength in Fig.\,\ref{fig_freqs1}. The symbols 
give the frequencies obtained with the Fourier transforms of the evolved
data and the solid lines are the corresponding linear fits, whose parameters are 
given in Table\,\ref{tab_fit_B1}. All Lower oscillations scale in the same way with 
the magnetic field strength such that $L_n=\frac{n}{2} L_2~~(n>2)$. For the Upper 
oscillations we find $U_n^\mathrm{core} = 3.7 n 
\,$Hz. The largest deviation from this scaling occurs for the two lowest QPOs,
$U_1^\mathrm{core}$ and $U_2^\mathrm{core}$, as we already reported in 
\cite{Gabler2012}. The frequencies of the higher overtones ($n>2$) are in
near integer relations $\frac{f_n}{f_m} \sim \frac{n}{m}$. 
Fig.~\ref{fig_freqs1} also shows that with increasing 
magnetic field strength the frequencies of the calculated QPOs deviate 
significantly from our linear fits towards lower values. This is expected 
since, as we  
described in \cite{Gabler2012},  the oscillations are no longer reflected efficiently 
from all parts of the core-crust interface as the field strength increases. 
Therefore, the oscillations move from the polar region towards the equatorial 
region, see e.g.~the changing location of $U_4^\mathrm{core}$ in 
Figs.\,\ref{fig_QPO_5_13} and \ref{fig_QPO_5_14}. Consequently, the 
corresponding frequencies in the continuum decrease. At magnetic field 
strengths $\bar B\gtrsim 
8\times10^{14}\,$G all Upper oscillations confined to the core disappear. They 
become coherent oscillations with constant phase, which can reach the 
surface. 

\begin{figure}
\includegraphics[width=.46\textwidth]{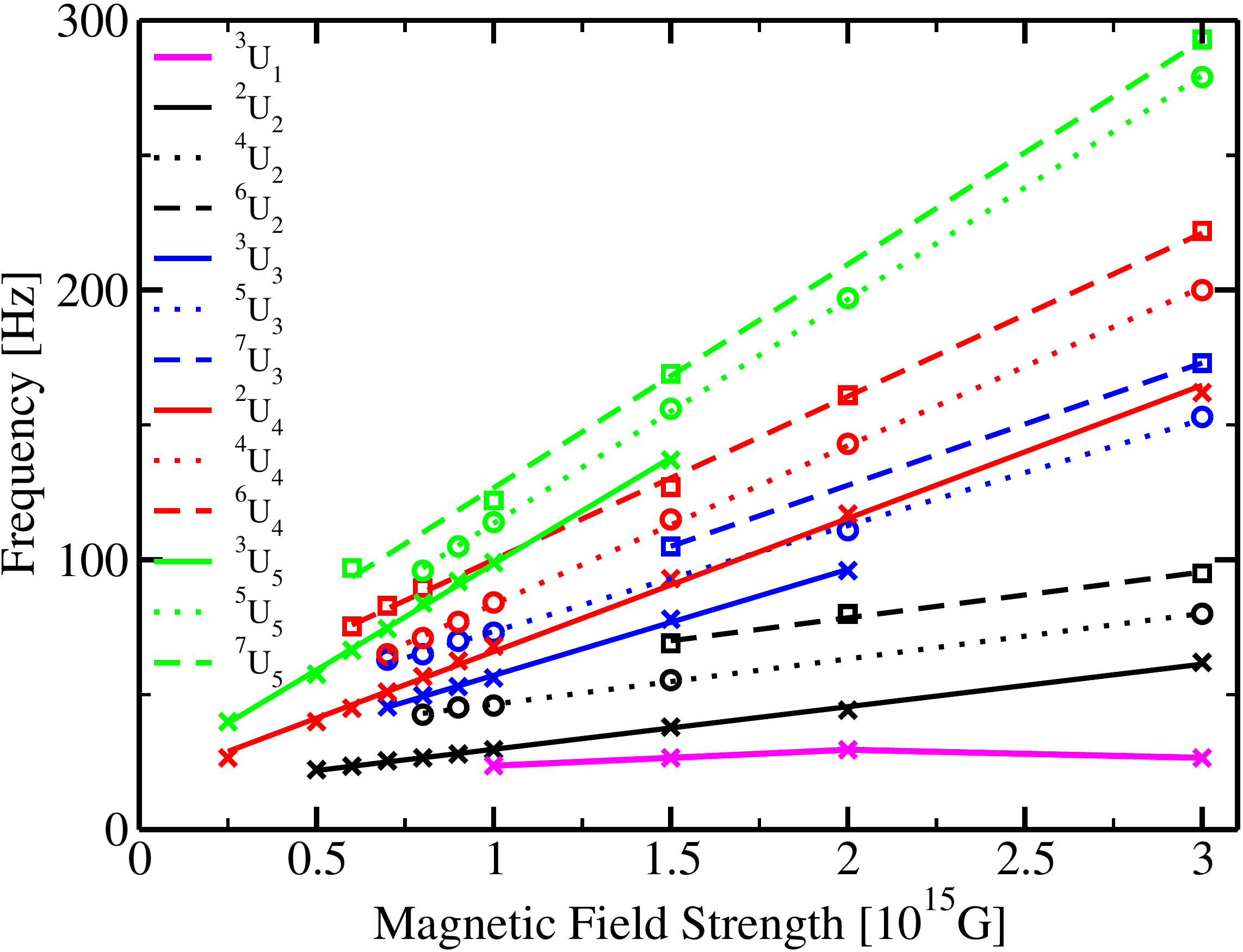}
\caption{Frequencies of the coherent oscillations having the lowest 
frequencies at as a function of magnetic field strength. Symbols give
the frequencies obtained from 
the simulations, and the lines are linear fits (see Table\,\ref{tab_fit_B2}).}
\label{fig_freqs2}
\end{figure}

\begin{table}
\begin{tabular}{c c c c c c c c c c c}
oscillation &$^2U_2$&$^4U_2$&$^6U_2$&$^3U_3$ &$^5U_3$&$^7U_3$\\ \hline 
$a$ [Hz]&14.0&29.6&44.5&17.8&34.2&37.0\\
$a_l^\mathrm{fit}$[Hz]&14.8&28.0&41.2&19.9&30.7&41.5\\
$b$ [Hz]&1.6&1.7&1.7&3.9&3.9&4.5\\
$b_n^\mathrm{fit}$ [Hz]&1.7&1.7&1.7&3.8&3.8&3.8\\
\hline\hline
\end{tabular}\\
\begin{tabular}{c c c c c c c c}
oscillation &$^2U_4$&$^4U_4$& $^6U_4$ &$^3U_5$ &$^5U_5$&$^7U_5$\\ \hline 
$a$ [Hz]&16.7&24.7&39.5&19.9&30.6&43.8\\ 
$a_l^\mathrm{fit}$[Hz]&14.8&28.0&41.2&19.9&30.7&41.5\\
$b$ [Hz]&4.9&5.9&6.0&7.9&8.3&8.3\\ 
$b_n^\mathrm{fit}$[Hz]&5.9&5.9&5.9&8.0&8.0&8.0\\ 
\hline
\end{tabular}
\caption{Linear fits of the frequency $f\,[\mathrm{Hz}] = a+ b \times \bar 
B[10^{14}\mathrm{G}]$ of the coherent oscillations of our superfluid 
model with 
realistic entrainment. The approximate values of these 
parameters obtained by assuming $a=a_l^\mathrm{fit}$ and 
$b=b_n^\mathrm{fit}$ are also shown. See main text for details.}
\label{tab_fit_B2}
\end{table}

The coherent oscillations also scale directly with the magnetic field 
strength (see Fig.\,\ref{fig_freqs2}, and the corresponding linear fits in 
Table\,\ref{tab_fit_B2}), which  
indicates that these oscillations cannot be crustal shear modes 
as the latter are independent of magnetic field strength. Furthermore, we 
see that oscillations having the same angular structure inside the 
crust, like $^2U_2$ and $^2U_4$, or $^4U_2$ and $^4U_4$, can have very 
different oscillation frequencies depending on the structure given by the 
number of nodes $n$ along the magnetic field lines. If these oscillations were 
crustal modes with no nodes in 
the radial direction inside the crust, they should have the same frequency only 
depending on $l$. The coherent oscillations also only appear for $\bar 
B\gtrsim2.5\times10^{14}\,$G, when the reflection of incoming oscillations at 
the core-crust boundary is no longer very strong. To fit the observed 
oscillations with a linear function $f=a+b\times\bar B[10^{14}\mathrm{G}]$, we 
also need to introduce an offset $a$. We prefer not to call the offset $a$ an 
asymptotic 
frequency for zero magnetic field strength, because for a vanishing magnetic 
field the coherent oscillations do not exist and only crustal modes 
survive. However, the presence of the offset $a$ also indicates that the 
coherent oscillations with constant phase are coupled, magneto-elastic 
oscillations that 
are not present without a solid crust. While the linear coefficient $b$ 
is nearly constant within 
one family of  coherent oscillations with the same $n$,
the offset $a$ increases with increasing $l$. For example, for $n=2$ and 
$l=\{2,3,4\}$ we find $a=\{14.0,29.6,44.5\}$Hz.
We can fit the factor $b$ reasonably well to $b=b_n^\mathrm{fit}\sim 
(2.1n-2.457)\,$Hz. The offset $a$ for a given $l$ is also similar for 
different values of $n$, i.e. we can approximate 
$a=a_l^\mathrm{fit}=(1.6+6.6 l)\,$Hz if $l$ is even, and 
$a=a_l^\mathrm{fit}=(3.7+5.4 l)\,$Hz if $l$ is odd. Therefore, the complete fits
read
\begin{eqnarray}
 f=\begin{cases}\label{eq_f_B_constant}
 (1.6+6.6 l + (2.1n-2.457)\bar B[10^{14}])\,\mathrm{Hz} &\text{$l$ even}\\
 (3.7+5.4 l + (2.1n-2.457)\bar B[10^{14}])\,\mathrm{Hz} &\text{$l$ odd}
   \end{cases}
\end{eqnarray}
With our assumptions that $a$ depends only on $l$, and $b$ depends only on $n$ 
we obtain reasonable fits to the measured values (Table\,\ref{tab_fit_B2}). The 
oscillation frequencies scale with the number of nodes $n$ along the magnetic 
field lines, indicating the Alfv\'enic character of the oscillations. Pure 
Alfv\'en oscillations behave in the same way. The offset scales with the angular 
number, similar to pure crustal shear modes. This similarity is intuitive, 
because for very low magnetic field strength crustal modes should appear. 
To the accuracy of the simulations
\begin{eqnarray}
 a=a_l^\mathrm{fit}\sim\sqrt{(l-1)(l+2)}, \label{eq_afit}
\end{eqnarray} 
as  it is the case for the purely crustal shear modes. However, given the 
numerical accuracy of the results we cannot distinguish between 
Eq.\,(\ref{eq_afit}) and the linear fit. Our fitting functions only allow us to 
understand the 
qualitative behaviour of the oscillations and not their actual behaviour  
since they are neither purely elastic nor purely magnetic.

\subsection{Numerical viscosity and different damping due to phase mixing}
\begin{figure}
\includegraphics[width=.46\textwidth]{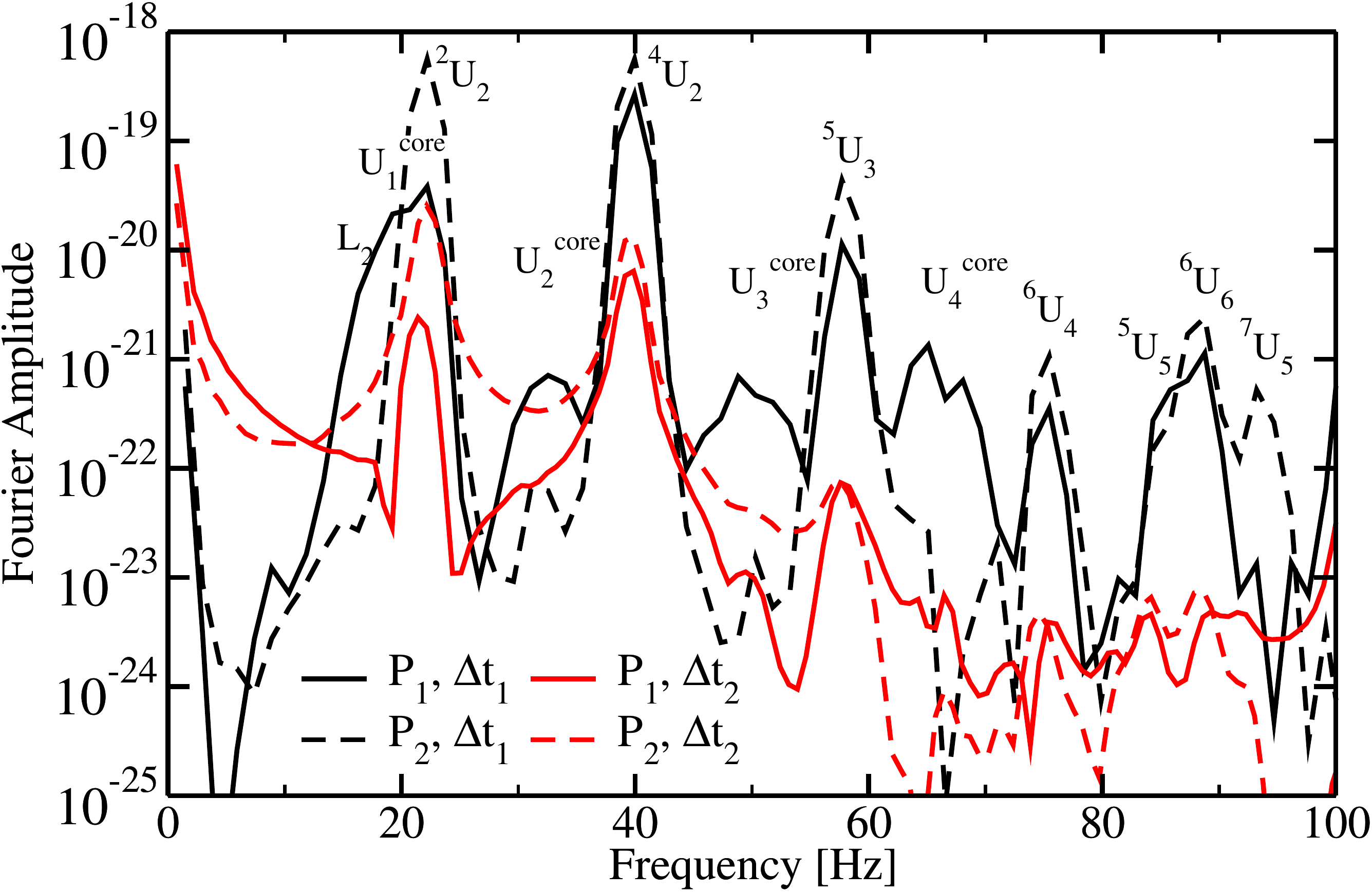}
\caption{Fourier amplitude of the velocity at points $P_1$ 
($\theta\sim1.2\,$rad, $r\sim3.5\,$km)  
in the core and $P_2$ ($\theta\sim0.8\,$rad, $r\sim9.2\,$km)
in the crust. Black lines are the results for the time interval $\Delta t_1=[0 
\,\mathrm{ms} ,675\,\mathrm{ms}]$ and red
lines for $\Delta t_2=[675\,\mathrm{ms},1350\,\mathrm{ms}]$.}
\label{fig_FFT_points}
\end{figure}
At $\bar B = 5\times10^{14}\,$G we find coexisting continuum and 
coherent, constant phase oscillations. However, we expect the  coherent
oscillations to persist for 
longer times, because the continuum oscillations have a salient damping 
mechanism in the form of phase mixing. 

In Fig.\,\ref{fig_FFT_points} we plot the 
Fourier amplitude of the velocity at two different locations inside the star, 
$P_1$ 
(in the core, $\theta\sim1.2\,$rad, $r\sim3.5\,$km) and 
$P_2$ (in the crust, $\theta\sim0.8\,$rad, $r\sim9.2\,$km), for different time 
intervals, 
$\Delta t_1=[0\,\mathrm{ms} ,675\,\mathrm{ms}]$ (black lines) and 
$\Delta t_2=[675\,\mathrm{ms},1350\,\mathrm{ms}]$ (red lines).
In the first interval, a number of coherent oscillations
($^2U_2$, $^4U_2$, $^5U_3$, $^6U_4$, $^5U_5$, $^6U_6$, and $^7U_5$) and 
continuum oscillations ($U_1^\mathrm{core}$, $U_2^\mathrm{core}$, 
$U_3^\mathrm{core}$, and $U_4^\mathrm{core}$) are excited initially.
Both families are damped by the numerical dissipation of the code and, in addition, the 
continuum oscillations undergo phase mixing. These damping mechanisms 
lead to a reduced amplitude of the coherent oscillations during the 
second 
interval $\Delta t_2$, while the continuum oscillations disappear 
completely. The finer the spatial structure of the coherent oscillations, 
the faster the damping \citep{Cerda2010}; as, for 
example, a comparison of the amplitudes of the oscillations 
$^4U_2$ and $^5U_3$ shows (Fig.\,\ref{fig_FFT_points}). 
The coherent oscillations with highest frequencies 
and finest spatial structures ($^6U_4$, $^5U_5$, $^6U_6$, and $^7U_5$) almost 
disappear completely due to numerical damping. 

\section{Parameter study of the entrainment effects}\label{sec_para}

We turn next to investigate how the results change when considering different 
values for the mass fraction that participates in the Alfv\'en motion of the 
core. Purely nucleonic EoS, like the APR EoS we use, consider 
neutrons, protons, electrons, and muons. These EoS predict proton
fractions of the order of $X_c\sim0.1-0.2$ \citep{Akmal1998} and 
effective masses that reduce this value to about $\varepsilon_\star X_c 
\sim0.05$ \citep{Chamel2008}. These values depend sensitively on the assumed 
composition in the neutron star core. For example, with the inclusion of hyperons (which are expected to appear from nuclear theory above a few times nuclear 
saturation density) the proton fraction may 
change to $X_c\sim0.3$ \citep{Baldo2000,Schulze2011}.

To investigate the effect on the oscillations of different compositions
in the neutron star core, we vary the factor $\varepsilon_\star X_c$ 
throughout the core, rescaling it by a 
constant number to encode our ignorance about the 
true EoS properties of the core.
In the following, the reported values of $\varepsilon_\star X_c$  
will always refer to its central value 
$\varepsilon_\star X_c(r=0)$, e.g. our reference value given in the APR EoS 
is $\varepsilon_\star^0 X^0_c(r=0)=0.0463$.
For our analysis we perform a new set of simulations 
varying $\varepsilon_\star X_c$ from its value inferred from theoretical 
calculation of the EoS, $\varepsilon_\star X_c =\varepsilon^0_\star X^0_c$, to 
the upper limit given by the normal fluid model,
 $\varepsilon_\star X_c=1.0$. The magnetic field strengths considered are 
$\bar B =\{0.5,1,2.5,5,7.5,10,15,20\}\times10^{14}\,$G, and the resolution 
employed is 
$150\times80$ zones for 
$r\times\theta=[0\,\text{km},10\,\text{km}]\times[0,\pi]$.

\begin{figure}
\includegraphics[width=.46\textwidth]{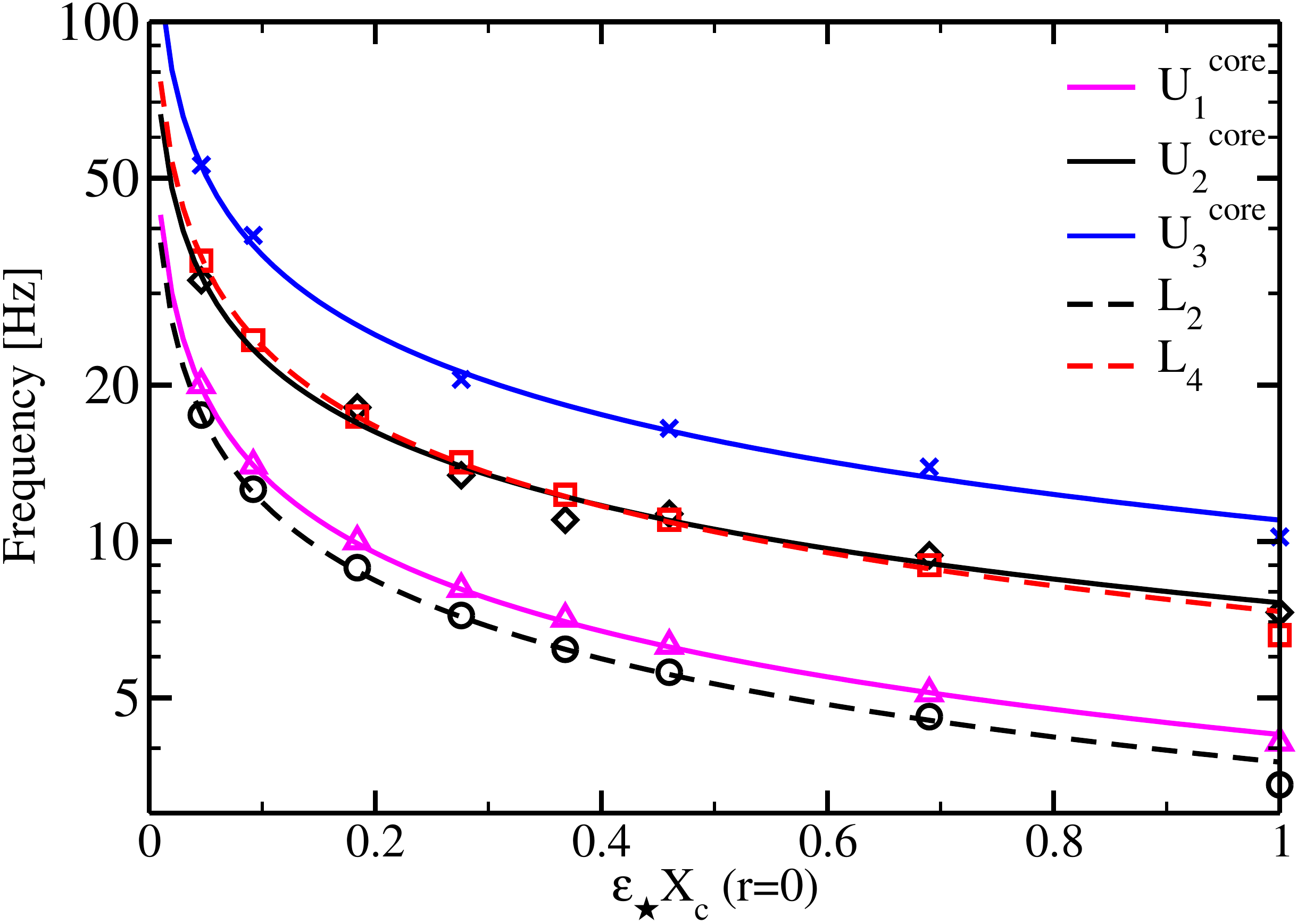}
\caption{Scaling of the frequencies with the parameters $\varepsilon_\star X_c$ 
for the first few Upper and Lower oscillations that are confined to the core, for 
$\bar B=5\times10^{14}\,$G. Lines are the fits with the parameters given in 
Table\,\ref{tab_fit_e1} and symbols are the results obtained in the 
simulations.}
\label{fig_core_e}
\end{figure}

From the propagation speeds (Eqs.\,(\ref{eq_EV_1}) and (\ref{eq_EV_2})) we 
expect that the Alfv\'en velocity at a given $\bar B$ in the core scales 
like 
\begin{eqnarray}
v_A(\varepsilon_\star 
X_c) \sim \frac{v_A(\varepsilon_\star X_c=1.0)}{\sqrt{\varepsilon_\star X_c}}\,.
\end{eqnarray}
Since the frequency of a pure Alfv\'en oscillation is proportional to the 
Alfv\'en speed, we expect the frequencies of the `core' oscillations 
($U_1^\mathrm{core},U_2^\mathrm{core},\cdots$) and Lower 
oscillations  ($L_2, L_4, \cdots$) to scale like 
\begin{eqnarray}
f_\mathrm{QPO}\sim d_0 \times 
\left(\varepsilon_\star X_c\right)^d\,, 
\end{eqnarray}
with $d=-0.5$. This assumption roughly agrees with our semi-analytic 
model to calculate the spectra in Fig.\,\ref{fig_spectrum}. The spectrum of the 
non-superfluid model can be scaled roughly as that of the 
superfluid model by dividing it by $\sqrt{\varepsilon^0_\star X^0_c}$. Our 
simulations at $\bar B=5\times10^{14}\,$G confirm this scaling approximately 
(see top two rows of Table\,\ref{tab_fit_e1}). 
In particular, the Lower oscillations scale as expected with exponents $d=-0.50$ 
and $d=-0.51$. (See also the fits in Fig.~\ref{fig_core_e}.) For the `core' 
oscillations the deviations are larger ($-0.59<d<-0.47$), which is 
probably 
related to the influence of the crust, where the reflection is not perfect.  
At $\bar{B}=5\times10^{14}\,$G oscillations already enter partially into the 
crust, and coherent oscillations appear. In this regime,  and for 
increasing magnetic 
field strength, the Upper oscillations move from the pole towards the equator, where 
the reflection at the core-crust interface is more efficient \citep{Gabler2012}. This spatial 
movement is accompanied by a shift in frequency, because the oscillation occurs at a 
different part of the continuum. This is visible in Fig.\,\ref{fig_spectrum} where for 
increasing crossing radius $\chi$ the frequency of the continuum decreases. 
A similar behaviour is expected by changing $\varepsilon_\star X_c$ instead of 
the magnetic field strength. The Upper oscillations can enter into the crust more 
efficiently for low $\varepsilon_\star X_c$, thus their maxima are shifted towards the 
equator and their frequencies shift to lower values in the spectrum. 
For this reason, the 
frequency of the core oscillations measured in the simulations does not scale 
exactly as $d\sim-0.5$. For very weak magnetic fields, $\bar B\lesssim5\times10^{13}\,$G
(where efficient reflection occurs and the Upper oscillations stay close to the polar axis) we 
thus expect a better agreement with $d=-0.5$. However, these are uninteresting 
magnetic field strengths for magnetars.

\begin{table}
\begin{tabular}{c c c c c c c c}
 & 
$L_2$&$L_4$&$U_1^\mathrm{core}$&$U_2^\mathrm{core}$&$U_3^\mathrm{core}
$&$U_4^\mathrm { core } $\\ \hline 
$d_0$& 3.76& 7.33&4.23&7.62&10.99&10.74\\
$d$&-0.50&-0.51&-0.50&-0.47&-0.51&-0.59\\
\hline\hline
\end{tabular}
\begin{tabular}{c c c c c c c c c c c}
 &$^2U_2$&$^4U_2$&$^1U_3$&$^3U_3$ &$^2U_4$&$^4U_4$&$^3U_5$\\ \hline 
$d_0$&16.2&16.7&23.3&35.1&40.4&50.9&55.0\\
$d$&-0.34&-0.58&-0.32&-0.33&-0.35&-0.34&-0.43\\
\hline\hline
\end{tabular}
\caption{Fittings of the frequencies $f\,[\mathrm{Hz}] = d_0 \times 
\left(\varepsilon_\star X_c\right)^d$ of various oscillations for $\bar 
B=5\times10^{14}\,$G (top) and 
$\bar 
B=2\times10^{15}\,$G (bottom).}
\label{tab_fit_e1}
\end{table}

\begin{figure}
\includegraphics[width=.46\textwidth]{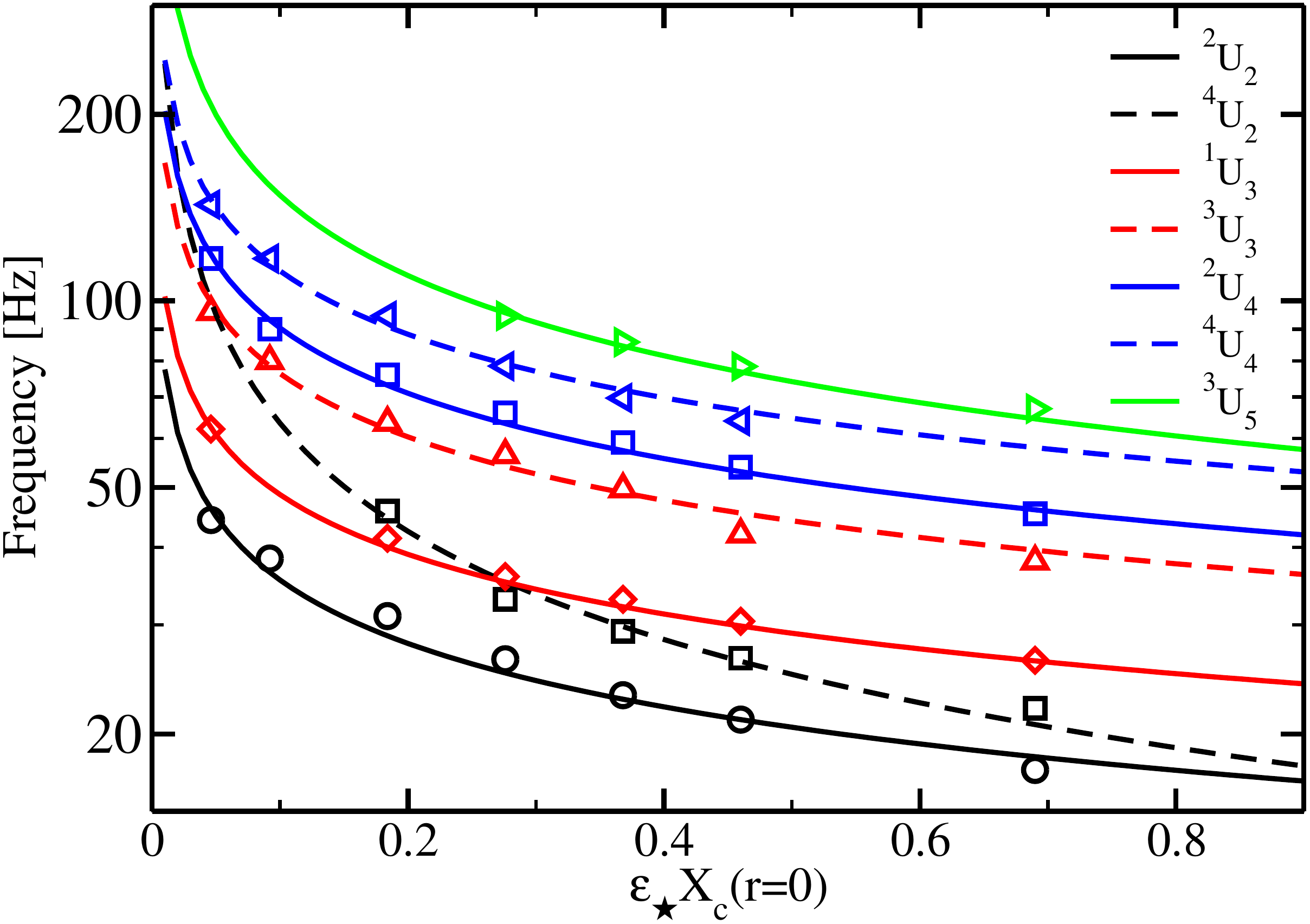}
\caption{Scaling of the frequencies with the parameters $\varepsilon_\star X_c$ 
for the first few coherent oscillations $^lU_n$, for $\bar 
B=2\times10^{15}\,$G. Lines are the fits with the parameters given in 
Table\,\ref{tab_fit_e1} and symbols are the results obtained in the simulations.
}
\label{fig_resonance_e}
\end{figure}

Regarding coherent oscillations (see the two bottom rows in 
Table\,\ref{tab_fit_e1} and 
Fig.\,\ref{fig_resonance_e}), their scaling with the entrainment factor and the 
proton fraction, 
$\left(\varepsilon_\star X_c\right)^d$ (which varies between $d=-0.58$ and 
$d=-0.32$, with most oscillations scaling as $d\sim-1/3$)  is not as close to 
$d=-0.5$ as for the continuum oscillations.
Nevertheless, this weaker scaling is expected because 
coherent oscillations have non-vanishing amplitudes inside the crust 
where the fraction of charged component and the entrainment are not changed 
for different simulations. We note that for the oscillations $^4U_2$ and 
$^3U_5$, which were the hardest to 
identify, we actually failed to identify them for very low values of
$\varepsilon_\star X_c (r=0)\lesssim0.3$. This may explain the larger 
deviations ($d=-0.58$ 
and $d=-0.43$) respectively, from the $d\sim-1/3$ scaling of the rest of 
oscillations.  Higher 
resolution and longer evolution times might reduce this discrepancy.

In the following, we proceed to find a general relation that describes the 
frequency of the dominant coherent oscillation $^2U_2$ for different 
magnetic field strengths and entrainment factors. The potential interest of
deriving such an equation is that it may help constrain the state of the matter 
inside the core of the neutron star by simply estimating the factor 
$\varepsilon_\star X_c$, if  the EoS, the stellar compactness, and the magnetic field
strength are known. To derive this equation we
make the following assumptions: (i) There is an offset $f^0_{^2U_2}$ which 
depends on $\varepsilon_\star X_c$, and may depend also on $l$ as we know 
from Eq.\,(\ref{eq_f_B_constant}), but does not depend on $\bar B$; 
(ii) 
the frequency is approximately proportional to $\bar B$, as we have seen in 
Section\,\ref{sec_scaling_B}; and (iii) the dependence on $\varepsilon_\star X_c$ 
can be described by a power law with exponent $b$. With these 
assumptions, we should find a relation of the form
\begin{eqnarray}\label{eq_fit_f}
 f_{^2U_2}[\mathrm{Hz}]=f^0_{^2U_2}(\varepsilon_\star X_c) + a_{^2U_2} \cdot
\left(\varepsilon_\star X_c\right)^{b_{^2U_2}}\cdot \bar B\,.
\end{eqnarray}

\begin{figure}
\includegraphics[width=.46\textwidth]{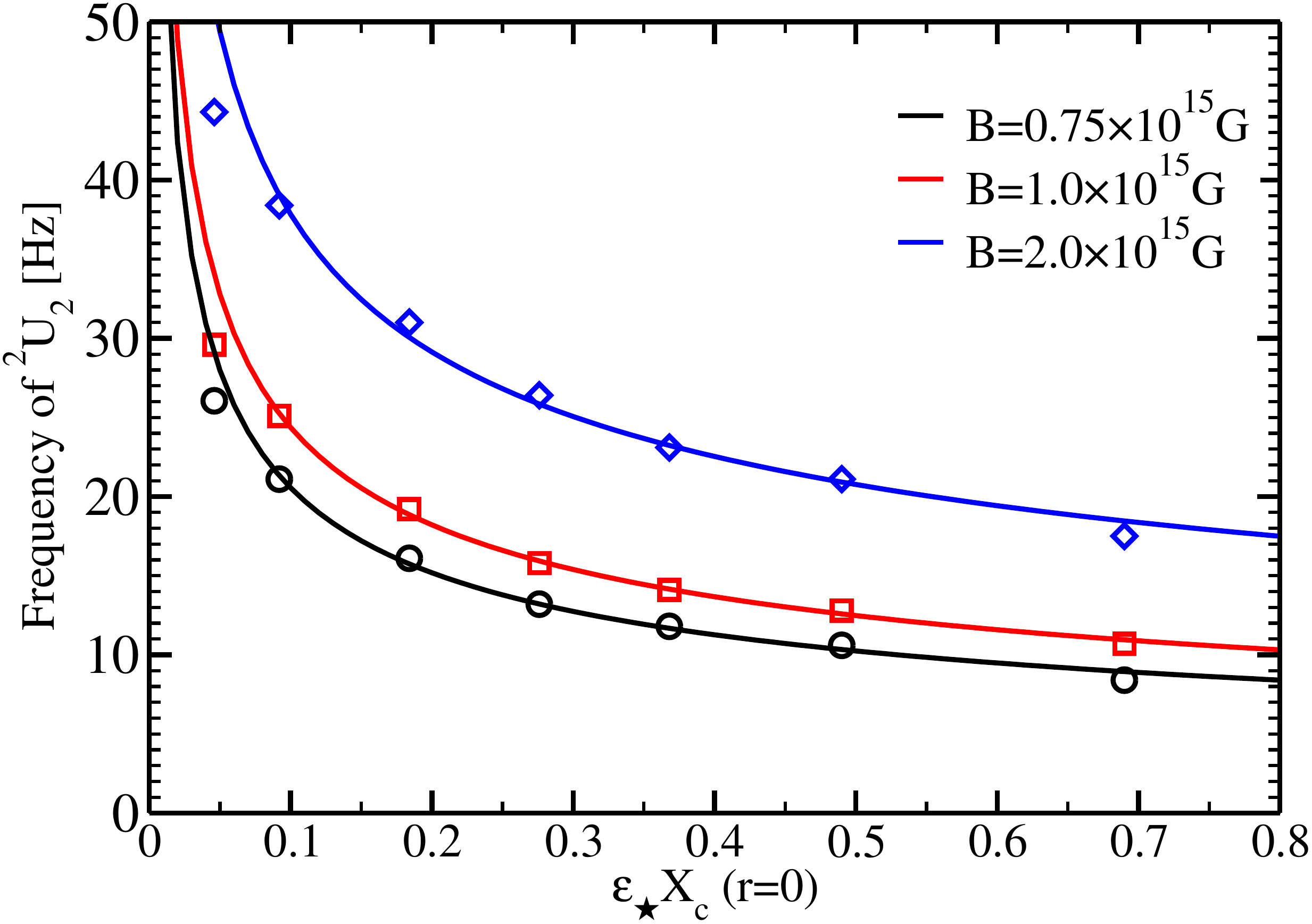}
\caption{Fitting functions defined in Eq.\,(\ref{eq_fit_f}) (solid lines) for 
three 
different magnetic field strengths, $\bar B=\{7.5,10,20\}\times10^{14}\,$G to 
obtain $f_{^2U_2}$ as a function of $\varepsilon_\star X_c$. Symbols give 
the frequencies obtained in the simulations.
}
\label{fig_f_eb}
\end{figure}

The first information we need is $f^0_{^2U_2}$ as a function of 
$\varepsilon_\star X_c$, which we can obtain from the 
dependence of the frequency on the magnetic field strength according to
\begin{eqnarray}\label{eq_fit_f_0}
 f\,[\mathrm{Hz}] = f^0_{^2U_2}(\varepsilon_\star X_c)+ a_{f^0} \times \bar 
B[10^{14}\mathrm{G}]\,.
\end{eqnarray}
Having found $f^0_{^2U_2}$ for different values of $\varepsilon_\star 
X_c$, we obtain a fitting function (see Table\,\ref{tab_fit_e2})
\begin{table}
\begin{tabular}{c c c c c c c c c c}
 $\varepsilon_\star X_c ~ [\varepsilon^0_\star 
X^0_c]$&1&2&4&6&8&10&15\\
$\varepsilon_\star X_c[10^{-2}]$&4.6&9.2&18.5&27.8&37.0&46.3&69.5\\
\hline 
$f^0_{^2U_2}\,$[Hz]& 14.9& 10.4&7.24&5.25&5.04&4.37&3.27\\
$a_{f^0}\,$[Hz]&1.47&1.41&1.19&1.06&0.90&0.84&0.71\\
\hline\hline
\end{tabular}
\caption{The frequency offset $f^0_{^2U_2}$ and the linear coefficient 
$a_{f^0}$ of the fitting functions 
Eq.\,(\ref{eq_fit_f_0}) of different entrainment factors for the 
oscillation $^2U_2$.}
\label{tab_fit_e2}
\end{table}
\begin{eqnarray}
 f^0_{^2U_2}=f^0_{^2U_2}(\varepsilon_\star 
X_c)=2.8\times(\varepsilon_\star X_c)^{-0.55}\,\mathrm{Hz}\,.
\end{eqnarray}
Next, we determine the parameters $a_{^2U_2}$ and $b_{^2U_2}$ in
Eq.\,(\ref{eq_fit_f}) by simulating the oscillations for different 
$\varepsilon_\star  X_c$ for three different magnetic field strengths,
$\bar B=\{7.5,10,20\}\times10^{14}\,$G. The resulting frequencies are
fitted by the parameters given in Table\,\ref{tab_fit_Be} and the corresponding 
curves are plotted in Fig.\,\ref{fig_f_eb}.
\begin{table}
\begin{tabular}{c | c c }
 $B[10^{14}G]$&$a_{^2U_2}$&$b_{^2U_2}$\\ \hline 
7.5&0.65&-0.34\\
10&0.66&-0.34\\
20&0.67&-0.32\\
\hline
$\varnothing$&0.66&-0.33\\
 \hline
 \hline
\end{tabular}
\caption{Parameters and their average for the fitting function
$f\,[\mathrm{Hz}] = f^0_{^2U_2}(\varepsilon_\star 
X_c) + a_{^2U_2} \cdot (\varepsilon_\star X_c)^{b_{^2U_2}}\cdot\bar 
B[10^{14}\mathrm{G}]$ for the coherent oscillation $^2U_2$ for 
three magnetic field strengths.}
\label{tab_fit_Be}
\end{table}
The parameters obtained for different magnetic field strengths agree well with 
each other, i.e. we can use their averages (see Table\,\ref{tab_fit_Be}) 
to finally obtain
\begin{eqnarray}\label{eq_fit_f_final}
  f_{^2U_2}[\mathrm{Hz}]= 2.8\times(\varepsilon_\star X_c)^{-0.55} + 
0.66 \left(\varepsilon_\star X_c\right)^{-0.33}\bar B 
[10^{14}\mathrm{G}] 
\,.
\end{eqnarray}

Fig.\ref{fig_f_eb} shows that the frequencies for the lowest values of 
$\varepsilon_\star X_c$ deviate most significantly from the fit  
(Eq.\,(\ref{eq_fit_f_final})). For such low values the crust, 
where the entrainment and the
fraction of charged particles is constant, sets a minimal timescale needed 
by the oscillations to cross the crust. This shortest travel time corresponds 
to the maximum frequency of the oscillations. As an extreme example, one could 
increase the propagation speed of the oscillation in the core to infinity by 
setting $\varepsilon_\star X_c=0$. However, in this case the maximum frequency 
is set by the crust alone (and is finite).

A final remark about Eq.\,(\ref{eq_fit_f_final}) is in order: coherent
oscillations with constant phase
exist not for all possible combinations of 
magnetic field strengths and choices of $\varepsilon_\star X_c$. For very high 
$\bar B$, Alfv\'en oscillations reach the surface of the magnetar while 
for very weak magnetic fields the oscillations are confined to the core. Likewise, the 
range of magnetic field strengths in which coherent oscillations can 
occur is wider for lower values of $\varepsilon_\star X_c$ and fairly narrow 
for non-superfluid matter $\varepsilon_\star X_c=1.0$ \citep{Gabler2013b}.

%
\section{Conclusions}\label{sec_discuss}

We have studied the effects of superfluidity on the magneto-elastic oscillations 
of magnetars for a realistic equation of state including stratification and entrainment. 
Our results confirm those of previous studies 
\citep{Gabler2013b,Passamonti2014} which  
showed the existence of both, continuous phase and coherent, constant 
phase oscillations
in the spectra of magnetars. 
In particular, we showed
that for typical magnetar field strengths ($\bar B \lesssim 5\times10^{15}\,$G) 
the continuous phase oscillations, also found in normal fluid models,  
are confined to the core of the neutron star, i.e.~they are Alfv\'en 
oscillations that are reflected at the core-crust interface and have negligible 
amplitudes inside the crust (see Fig.\,\ref{fig_QPO_5_13}). 

Coherent oscillations, with properties very different from 
continuous phase oscillations, are a consequence of the relatively stronger 
coupling of oscillations through the crust in superfluid models compared to 
their normal fluid counterparts. 
Our simulations have shown that coherent oscillations have the following 
properties: (i) They have non-vanishing amplitudes inside the crust 
(see Figs.\,\ref{fig_QPO_5_14} and \ref{fig_QPO_2_15}). (ii) They appear (depending 
on the proton fraction and the entrainment) at $\bar 
B\gtrsim\mathrm{few}\times10^{14}\,G$ (see Fig.\,\ref{fig_freqs2}). (iii)  
The increased coupling of different field lines through the crust allows these 
oscillations to have structures perpendicular to the field lines. As a result,
an oscillation, which is characterised by the number of maxima $n$ along  
field lines only, splits now into several oscillations with different numbers 
of maxima $l$ perpendicular to the field lines (see Figs.\,\ref{fig_QPO_5_14} 
and 
\ref{fig_QPO_2_15}). (iv) Due to the constant phase, the newly found 
oscillations are longer-lived (see Fig.\,\ref{fig_FFT_points}).  (v) Their 
frequencies  scale linearly with the magnetic field strength, but have a constant 
offset. They are thus clearly neither pure Alfv\'en oscillation nor crustal shear modes. 
These $^lU_n$ oscillations are the most promising candidates for explaining the 
observed low frequency, $f\lesssim155\,$Hz, QPOs in magnetars.

We use the term coherent oscillations instead of discrete normal 
modes, because the latter would imply that the oscillations are orthogonal 
and form a complete set of eigenmodes, which we cannot proof rigorously
with our direct numerical simulations. Performing a normal mode analysis it may 
be possible to show that discrete normal modes indeed exist. 
\citet{Sotani2015} found normal modes in magnetar models with highly 
tangled magnetic fields. However, these eigenmodes are different from our 
 coherent oscillations with constant phase, because they do not have a 
frequency offset of the order of the crustal shear modes and they do exist at 
all magnetic field strengths, i.e. there is no continuum in the limit of weak or 
strong magnetic fields. However, the strong coupling due to the high 
entanglement may play a similar role as the strong coupling of different field 
lines through the crust in this work.

In our study, the effects of changing the composition and entrainment were 
encoded into a single parameter, $\varepsilon_\star X_c$. As 
expected, the frequencies of the oscillations confined to the core, 
$U^\mathrm{core}_n$ and $L_n$, scale like $f\sim d_0\sqrt{\varepsilon_\star 
X_c}$. Coherent oscillations, on the contrary, have a different scaling, 
$f\sim f^0 (\varepsilon_\star X_c) + a_{f^0} \left(\varepsilon_\star X_c\right)^{-1/3} 
\bar B$, where the exponent is $-1/3$ for several $^lU_n$ 
oscillations for a given magnetic field strength. In the particular case of  
$^2U_2$, it was obtained for three different magnetic field strengths. 

The superfluid magneto-elastic oscillations discussed in this work (see also 
\cite{Gabler2013b,Passamonti2014}) allow to explain qualitatively, and 
simultaneously, both the low- and high-frequency QPOs  observed in magnetars. 
However, our model still has to be improved. On the one hand, it
 does not include the possibility of superconducting protons (which may be destroyed due to the ultra-strong magnetic fields present in magnetars). On the 
other hand, we do neither consider a highly tangled magnetic field nor 
mixed toroidal and poloidal magnetic fields that may change the 
spectrum \citep{Colaiuda2012, Sotani2015}.
We plan to account for these two main 
missing ingredients in the near future and address these issues in further work. 
Another important open question is how the oscillations of a neutron star may 
modulate its X-ray emission. To understand the modulation mechanism it would be 
necessary to estimate the amplitude of the oscillations and to be able 
to compare them with the properties of observed QPOs. Needless to say, to {\it 
really} advance in our understanding of  QPOs, we eagerly need to increase the 
available observational data. There is still no unambiguous pattern in the 
frequency spacing of the QPOs. Recent work aimed at detecting QPOs in normal 
magnetar bursts appears very promising 
\citep{Huppenkothen2014a,Huppenkothen2014b} and we strongly encourage further 
studies in this direction.

\section*{Acknowledgements}

Work supported by the Collaborative Research Centre on Gravitational
Wave Astronomy of the Deutsche Forschungsgemeinschaft (DFG
SFB/Transregio 7), the Spanish MINECO (grant AYA2013-40979-P),
 the {\it Generalitat Valenciana} (PROMETEOII-2014-069), and the EU through the ERC
Starting Grant no. 259276-CAMAP and the ERC Advanced Grant
no. 341157-COCO2CASA. Partial support comes from NewCompStar, COST
Action MP1304. Computations were performed at the {\it Servei
  d'Inform\`atica de la Universitat de Val\`encia} and at the 
Max Planck Computing and Data Facility (MPCDF).

\bibliographystyle{mnras}
\bibliography{magnetar}

\begin{thebibliography}{}
\makeatletter
\relax
\def\mn@urlcharsother{\let\do\@makeother \do\$\do\&\do\#\do\^\do\_\do\%\do\~}
\def\mn@doi{\begingroup\mn@urlcharsother \@ifnextchar [ {\mn@doi@}
  {\mn@doi@[]}}
\def\mn@doi@[#1]#2{\def\@tempa{#1}\ifx\@tempa\@empty \href
  {http://dx.doi.org/#2} {doi:#2}\else \href {http://dx.doi.org/#2} {#1}\fi
  \endgroup}
\def\mn@eprint#1#2{\mn@eprint@#1:#2::\@nil}
\def\mn@eprint@arXiv#1{\href {http://arxiv.org/abs/#1} {{\tt arXiv:#1}}}
\def\mn@eprint@dblp#1{\href {http://dblp.uni-trier.de/rec/bibtex/#1.xml}
  {dblp:#1}}
\def\mn@eprint@#1:#2:#3:#4\@nil{\def\@tempa {#1}\def\@tempb {#2}\def\@tempc
  {#3}\ifx \@tempc \@empty \let \@tempc \@tempb \let \@tempb \@tempa \fi \ifx
  \@tempb \@empty \def\@tempb {arXiv}\fi \@ifundefined
  {mn@eprint@\@tempb}{\@tempb:\@tempc}{\expandafter \expandafter \csname
  mn@eprint@\@tempb\endcsname \expandafter{\@tempc}}}

\bibitem[\protect\citeauthoryear{{Akmal}, {Pandharipande}  \&
  {Ravenhall}}{{Akmal} et~al.}{1998}]{Akmal1998}
{Akmal} A.,  {Pandharipande} V.~R.,   {Ravenhall} D.~G.,  1998, \mn@doi [\prc]
  {10.1103/PhysRevC.58.1804}, \href
  {http://adsabs.harvard.edu/abs/1998PhRvC..58.1804A} {58, 1804}

\bibitem[\protect\citeauthoryear{{Anderson} \& {Itoh}}{{Anderson} \&
  {Itoh}}{1975}]{Anderson1975}
{Anderson} P.~W.,  {Itoh} N.,  1975, \mn@doi [\nat] {10.1038/256025a0}, \href
  {http://adsabs.harvard.edu/abs/1975Natur.256...25A} {256, 25}

\bibitem[\protect\citeauthoryear{{Andersson}, {Comer}  \&
  {Langlois}}{{Andersson} et~al.}{2002}]{Andersson2002}
{Andersson} N.,  {Comer} G.~L.,   {Langlois} D.,  2002, \mn@doi [\prd]
  {10.1103/PhysRevD.66.104002}, \href
  {http://adsabs.harvard.edu/abs/2002PhRvD..66j4002A} {66, 104002}

\bibitem[\protect\citeauthoryear{{Andersson}, {Comer}  \&
  {Grosart}}{{Andersson} et~al.}{2004}]{Andersson2004}
{Andersson} N.,  {Comer} G.~L.,   {Grosart} K.,  2004, \mn@doi [\mnras]
  {10.1111/j.1365-2966.2004.08370.x}, \href
  {http://adsabs.harvard.edu/abs/2004MNRAS.355..918A} {355, 918}

\bibitem[\protect\citeauthoryear{{Andersson}, {Glampedakis}  \&
  {Samuelsson}}{{Andersson} et~al.}{2009}]{Andersson2009}
{Andersson} N.,  {Glampedakis} K.,   {Samuelsson} L.,  2009, \mn@doi [\mnras]
  {10.1111/j.1365-2966.2009.14734.x}, \href
  {http://adsabs.harvard.edu/abs/2009MNRAS.396..894A} {396, 894}

\bibitem[\protect\citeauthoryear{{Baldo}, {Burgio}  \& {Schulze}}{{Baldo}
  et~al.}{2000}]{Baldo2000}
{Baldo} M.,  {Burgio} G.~F.,   {Schulze} H.-J.,  2000, \mn@doi [\prc]
  {10.1103/PhysRevC.61.055801}, \href
  {http://adsabs.harvard.edu/abs/2000PhRvC..61e5801B} {61, 055801}

\bibitem[\protect\citeauthoryear{{Baym}, {Pethick}  \& {Pines}}{{Baym}
  et~al.}{1969}]{Baym1969}
{Baym} G.,  {Pethick} C.,   {Pines} D.,  1969, \mn@doi [\nat]
  {10.1038/224673a0}, \href {http://adsabs.harvard.edu/abs/1969Natur.224..673B}
  {224, 673}

\bibitem[\protect\citeauthoryear{{Bocquet}, {Bonazzola}, {Gourgoulhon}  \&
  {Novak}}{{Bocquet} et~al.}{1995}]{Bocquet1995}
{Bocquet} M.,  {Bonazzola} S.,  {Gourgoulhon} E.,   {Novak} J.,  1995, \aap,
  \href {http://adsabs.harvard.edu/abs/1995A%26A...301..757B} {301, 757}

\bibitem[\protect\citeauthoryear{{Cerd{\'a}-Dur{\'a}n}}{{Cerd{\'a}-Dur{\'a}n}}{2010}]{Cerda2010}
{Cerd{\'a}-Dur{\'a}n} P.,  2010, \mn@doi [Classical and Quantum Gravity]
  {10.1088/0264-9381/27/20/205012}, \href
  {http://adsabs.harvard.edu/abs/2010CQGra..27t5012C} {27, 205012}

\bibitem[\protect\citeauthoryear{{Cerd{\'a}-Dur{\'a}n}, {Font}, {Ant{\'o}n}  \&
  {M{\"u}ller}}{{Cerd{\'a}-Dur{\'a}n} et~al.}{2008}]{Cerda2008}
{Cerd{\'a}-Dur{\'a}n} P.,  {Font} J.~A.,  {Ant{\'o}n} L.,   {M{\"u}ller} E.,
  2008, \mn@doi [\aap] {10.1051/0004-6361:200810086}, \href
  {http://adsabs.harvard.edu/abs/2008A%26A...492..937C} {492, 937}

\bibitem[\protect\citeauthoryear{{Cerd{\'a}-Dur{\'a}n}, {Stergioulas}  \&
  {Font}}{{Cerd{\'a}-Dur{\'a}n} et~al.}{2009}]{Cerda2009}
{Cerd{\'a}-Dur{\'a}n} P.,  {Stergioulas} N.,   {Font} J.~A.,  2009, \mn@doi
  [\mnras] {10.1111/j.1365-2966.2009.15056.x}, \href
  {http://adsabs.harvard.edu/abs/2009MNRAS.397.1607C} {397, 1607}

\bibitem[\protect\citeauthoryear{{Chamel}}{{Chamel}}{2008}]{Chamel2008b}
{Chamel} N.,  2008, \mn@doi [\mnras] {10.1111/j.1365-2966.2008.13426.x}, \href
  {http://adsabs.harvard.edu/abs/2008MNRAS.388..737C} {388, 737}

\bibitem[\protect\citeauthoryear{{Chamel}}{{Chamel}}{2012}]{Chamel2012}
{Chamel} N.,  2012, \mn@doi [\prc] {10.1103/PhysRevC.85.035801}, \href
  {http://adsabs.harvard.edu/abs/2012PhRvC..85c5801C} {85, 035801}

\bibitem[\protect\citeauthoryear{Chamel \& Haensel}{Chamel \&
  Haensel}{2008}]{Chamel2008}
Chamel N.,  Haensel P.,  2008, Living Reviews in Relativity, 11

\bibitem[\protect\citeauthoryear{{Colaiuda} \& {Kokkotas}}{{Colaiuda} \&
  {Kokkotas}}{2011}]{Colaiuda2011}
{Colaiuda} A.,  {Kokkotas} K.~D.,  2011, \mn@doi [\mnras]
  {10.1111/j.1365-2966.2011.18602.x}, \href
  {http://adsabs.harvard.edu/abs/2011MNRAS.414.3014C} {414, 3014}

\bibitem[\protect\citeauthoryear{{Colaiuda} \& {Kokkotas}}{{Colaiuda} \&
  {Kokkotas}}{2012}]{Colaiuda2012}
{Colaiuda} A.,  {Kokkotas} K.~D.,  2012, \mn@doi [\mnras]
  {10.1111/j.1365-2966.2012.20919.x}, \href
  {http://adsabs.harvard.edu/abs/2012MNRAS.423..811C} {423, 811}

\bibitem[\protect\citeauthoryear{{Colaiuda}, {Beyer}  \& {Kokkotas}}{{Colaiuda}
  et~al.}{2009}]{Colaiuda2009}
{Colaiuda} A.,  {Beyer} H.,   {Kokkotas} K.~D.,  2009, \mn@doi [\mnras]
  {10.1111/j.1365-2966.2009.14878.x}, \href
  {http://adsabs.harvard.edu/abs/2009MNRAS.396.1441C} {396, 1441}

\bibitem[\protect\citeauthoryear{{Deibel}, {Steiner}  \& {Brown}}{{Deibel}
  et~al.}{2014}]{Deibel2014}
{Deibel} A.~T.,  {Steiner} A.~W.,   {Brown} E.~F.,  2014, \mn@doi [\prc]
  {10.1103/PhysRevC.90.025802}, \href
  {http://adsabs.harvard.edu/abs/2014PhRvC..90b5802D} {90, 025802}

\bibitem[\protect\citeauthoryear{{Douchin} \& {Haensel}}{{Douchin} \&
  {Haensel}}{2001}]{Douchin2001}
{Douchin} F.,  {Haensel} P.,  2001, \mn@doi [\aap]
  {10.1051/0004-6361:20011402}, \href
  {http://adsabs.harvard.edu/abs/2001A%26A...380..151D} {380, 151}

\bibitem[\protect\citeauthoryear{{Duncan}}{{Duncan}}{1998}]{Duncan1998}
{Duncan} R.~C.,  1998, \mn@doi [\apjl] {10.1086/311303}, \href
  {http://adsabs.harvard.edu/abs/1998ApJ...498L..45D} {498, L45}

\bibitem[\protect\citeauthoryear{{Duncan} \& {Thompson}}{{Duncan} \&
  {Thompson}}{1992}]{Duncan1992}
{Duncan} R.~C.,  {Thompson} C.,  1992, \mn@doi [\apjl] {10.1086/186413}, \href
  {http://adsabs.harvard.edu/abs/1992ApJ...392L...9D} {392, L9}

\bibitem[\protect\citeauthoryear{{Gabler}, {Cerd{\'a} Dur{\'a}n}, {Font},
  {M{\"u}ller}  \& {Stergioulas}}{{Gabler} et~al.}{2011}]{Gabler2011letter}
{Gabler} M.,  {Cerd{\'a} Dur{\'a}n} P.,  {Font} J.~A.,  {M{\"u}ller} E.,
  {Stergioulas} N.,  2011, \mn@doi [\mnras] {10.1111/j.1745-3933.2010.00974.x},
  \href {http://adsabs.harvard.edu/abs/2011MNRAS.410L..37G} {410, L37}

\bibitem[\protect\citeauthoryear{{Gabler}, {Cerd{\'a}-Dur{\'a}n},
  {Stergioulas}, {Font}  \& {M{\"u}ller}}{{Gabler} et~al.}{2012}]{Gabler2012}
{Gabler} M.,  {Cerd{\'a}-Dur{\'a}n} P.,  {Stergioulas} N.,  {Font} J.~A.,
  {M{\"u}ller} E.,  2012, \mn@doi [\mnras] {10.1111/j.1365-2966.2012.20454.x},
  \href {http://adsabs.harvard.edu/abs/2012MNRAS.421.2054G} {421, 2054}

\bibitem[\protect\citeauthoryear{{Gabler}, {Cerd{\'a}-Dur{\'a}n},
  {Stergioulas}, {Font}  \& {M{\"u}ller}}{{Gabler} et~al.}{2013a}]{Gabler2013b}
{Gabler} M.,  {Cerd{\'a}-Dur{\'a}n} P.,  {Stergioulas} N.,  {Font} J.~A.,
  {M{\"u}ller} E.,  2013a, \mn@doi [Physical Review Letters]
  {10.1103/PhysRevLett.111.211102}, \href
  {http://adsabs.harvard.edu/abs/2013PhRvL.111u1102G} {111, 211102}

\bibitem[\protect\citeauthoryear{{Gabler}, {Cerd{\'a}-Dur{\'a}n}, {Font},
  {M{\"u}ller}  \& {Stergioulas}}{{Gabler} et~al.}{2013b}]{Gabler2013a}
{Gabler} M.,  {Cerd{\'a}-Dur{\'a}n} P.,  {Font} J.~A.,  {M{\"u}ller} E.,
  {Stergioulas} N.,  2013b, \mn@doi [\mnras] {10.1093/mnras/sts721}, \href
  {http://adsabs.harvard.edu/abs/2013MNRAS.430.1811G} {430, 1811}

\bibitem[\protect\citeauthoryear{{Gabler}, {Cerd{\'a}-Dur{\'a}n},
  {Stergioulas}, {Font}  \& {M{\"u}ller}}{{Gabler} et~al.}{2014}]{Gabler2014b}
{Gabler} M.,  {Cerd{\'a}-Dur{\'a}n} P.,  {Stergioulas} N.,  {Font} J.~A.,
  {M{\"u}ller} E.,  2014, \mn@doi [\mnras] {10.1093/mnras/stu1263}, \href
  {http://adsabs.harvard.edu/abs/2014MNRAS.443.1416G} {443, 1416}

\bibitem[\protect\citeauthoryear{{Gearheart}, {Newton}, {Hooker}  \&
  {Li}}{{Gearheart} et~al.}{2011}]{Gearheart2011}
{Gearheart} M.,  {Newton} W.~G.,  {Hooker} J.,   {Li} B.-A.,  2011, \mn@doi
  [\mnras] {10.1111/j.1365-2966.2011.19628.x}, \href
  {http://adsabs.harvard.edu/abs/2011MNRAS.418.2343G} {418, 2343}

\bibitem[\protect\citeauthoryear{{Glampedakis} \& {Jones}}{{Glampedakis} \&
  {Jones}}{2014}]{Glampedakis2014b}
{Glampedakis} K.,  {Jones} D.~I.,  2014, \mn@doi [\mnras]
  {10.1093/mnras/stu017}, \href
  {http://adsabs.harvard.edu/abs/2014MNRAS.439.1522G} {439, 1522}

\bibitem[\protect\citeauthoryear{{Glampedakis}, {Samuelsson}  \&
  {Andersson}}{{Glampedakis} et~al.}{2006}]{Glampedakis2006b}
{Glampedakis} K.,  {Samuelsson} L.,   {Andersson} N.,  2006, \mn@doi [\mnras]
  {10.1111/j.1745-3933.2006.00211.x}, \href
  {http://adsabs.harvard.edu/abs/2006MNRAS.371L..74G} {371, L74}

\bibitem[\protect\citeauthoryear{{Glampedakis}, {Andersson}  \&
  {Samuelsson}}{{Glampedakis} et~al.}{2011}]{Glampedakis2011a}
{Glampedakis} K.,  {Andersson} N.,   {Samuelsson} L.,  2011, \mn@doi [\mnras]
  {10.1111/j.1365-2966.2010.17484.x}, \href
  {http://adsabs.harvard.edu/abs/2011MNRAS.410..805G} {410, 805}

\bibitem[\protect\citeauthoryear{{Graber}, {Andersson}, {Glampedakis}  \&
  {Lander}}{{Graber} et~al.}{2015}]{Graber2015}
{Graber} V.,  {Andersson} N.,  {Glampedakis} K.,   {Lander} S.~K.,  2015,
  \mn@doi [\mnras] {10.1093/mnras/stv1648}, \href
  {http://adsabs.harvard.edu/abs/2015MNRAS.453..671G} {453, 671}

\bibitem[\protect\citeauthoryear{{Gusakov} \& {Haensel}}{{Gusakov} \&
  {Haensel}}{2005}]{Gusakov2005}
{Gusakov} M.~E.,  {Haensel} P.,  2005, \mn@doi [Nuclear Physics A]
  {10.1016/j.nuclphysa.2005.07.005}, \href
  {http://adsabs.harvard.edu/abs/2005NuPhA.761..333G} {761, 333}

\bibitem[\protect\citeauthoryear{{Hambaryan}, {Neuh{\"a}user}  \&
  {Kokkotas}}{{Hambaryan} et~al.}{2011}]{Hambaryan2011}
{Hambaryan} V.,  {Neuh{\"a}user} R.,   {Kokkotas} K.~D.,  2011, \mn@doi [\aap]
  {10.1051/0004-6361/201015273}, \href
  {http://adsabs.harvard.edu/abs/2011A%26A...528A..45H} {528, A45+}

\bibitem[\protect\citeauthoryear{{Huppenkothen} et~al.,}{{Huppenkothen}
  et~al.}{2014a}]{Huppenkothen2014a}
{Huppenkothen} D.,  et~al., 2014a, \mn@doi [\apj]
  {10.1088/0004-637X/787/2/128}, \href
  {http://adsabs.harvard.edu/abs/2014ApJ...787..128H} {787, 128}

\bibitem[\protect\citeauthoryear{{Huppenkothen}, {Watts}  \&
  {Levin}}{{Huppenkothen} et~al.}{2014b}]{Huppenkothen2014c}
{Huppenkothen} D.,  {Watts} A.~L.,   {Levin} Y.,  2014b, \mn@doi [\apj]
  {10.1088/0004-637X/793/2/129}, \href
  {http://adsabs.harvard.edu/abs/2014ApJ...793..129H} {793, 129}

\bibitem[\protect\citeauthoryear{{Huppenkothen}, {Heil}, {Watts}  \& {G{\"o}{\u
  g}{\"u}{\c s}}}{{Huppenkothen} et~al.}{2014c}]{Huppenkothen2014b}
{Huppenkothen} D.,  {Heil} L.~M.,  {Watts} A.~L.,   {G{\"o}{\u g}{\"u}{\c s}}
  E.,  2014c, \mn@doi [\apj] {10.1088/0004-637X/795/2/114}, \href
  {http://adsabs.harvard.edu/abs/2014ApJ...795..114H} {795, 114}

\bibitem[\protect\citeauthoryear{{Israel} et~al.,}{{Israel}
  et~al.}{2005}]{Israel2005}
{Israel} G.~L.,  et~al., 2005, \mn@doi [\apjl] {10.1086/432615}, \href
  {http://adsabs.harvard.edu/abs/2005ApJ...628L..53I} {628, L53}

\bibitem[\protect\citeauthoryear{{Levin}}{{Levin}}{2006}]{Levin2006}
{Levin} Y.,  2006, \mn@doi [\mnras] {10.1111/j.1745-3933.2006.00155.x}, \href
  {http://adsabs.harvard.edu/abs/2006MNRAS.368L..35L} {368, L35}

\bibitem[\protect\citeauthoryear{{Levin}}{{Levin}}{2007}]{Levin2007}
{Levin} Y.,  2007, \mn@doi [\mnras] {10.1111/j.1365-2966.2007.11582.x}, \href
  {http://adsabs.harvard.edu/abs/2007MNRAS.377..159L} {377, 159}

\bibitem[\protect\citeauthoryear{{Link}}{{Link}}{2014}]{Link2014}
{Link} B.,  2014, \mn@doi [\mnras] {10.1093/mnras/stu584}, \href
  {http://adsabs.harvard.edu/abs/2014MNRAS.441.2676L} {441, 2676}

\bibitem[\protect\citeauthoryear{{Link} \& {van Eysden}}{{Link} \& {van
  Eysden}}{2015}]{Link2015}
{Link} B.,  {van Eysden} C.~A.,  2015, preprint, \href
  {http://adsabs.harvard.edu/abs/2015arXiv150301410L} {} (\mn@eprint {arXiv}
  {1503.01410})

\bibitem[\protect\citeauthoryear{{Mendell}}{{Mendell}}{1991}]{Mendell1991}
{Mendell} G.,  1991, \mn@doi [\apj] {10.1086/170609}, \href
  {http://adsabs.harvard.edu/abs/1991ApJ...380..515M} {380, 515}

\bibitem[\protect\citeauthoryear{{Mendell}}{{Mendell}}{1998}]{Mendell1998}
{Mendell} G.,  1998, \mn@doi [\mnras] {10.1046/j.1365-8711.1998.01451.x}, \href
  {http://adsabs.harvard.edu/abs/1998MNRAS.296..903M} {296, 903}

\bibitem[\protect\citeauthoryear{{Messios}, {Papadopoulos}  \&
  {Stergioulas}}{{Messios} et~al.}{2001}]{Messios2001}
{Messios} N.,  {Papadopoulos} D.~B.,   {Stergioulas} N.,  2001, \mn@doi
  [\mnras] {10.1046/j.1365-8711.2001.04645.x}, \href
  {http://adsabs.harvard.edu/abs/2001MNRAS.328.1161M} {328, 1161}

\bibitem[\protect\citeauthoryear{{Migdal}}{{Migdal}}{1959}]{Migdal1959}
{Migdal} A.~B.,  1959, \mn@doi [\nphysa] {10.1016/0029-5582(59)90264-0}, \href
  {http://adsabs.harvard.edu/abs/1959NucPh..13..655M} {13, 655}

\bibitem[\protect\citeauthoryear{{Olausen} \& {Kaspi}}{{Olausen} \&
  {Kaspi}}{2013}]{Olausen2013}
{Olausen} S.~A.,  {Kaspi} V.~M.,  2013, preprint, \href
  {http://adsabs.harvard.edu/abs/2013arXiv1309.4167O} {} (\mn@eprint {arXiv}
  {1309.4167})

\bibitem[\protect\citeauthoryear{{Page}, {Prakash}, {Lattimer}  \&
  {Steiner}}{{Page} et~al.}{2011}]{Page2011}
{Page} D.,  {Prakash} M.,  {Lattimer} J.~M.,   {Steiner} A.~W.,  2011, \mn@doi
  [Physical Review Letters] {10.1103/PhysRevLett.106.081101}, \href
  {http://adsabs.harvard.edu/abs/2011PhRvL.106h1101P} {106, 081101}

\bibitem[\protect\citeauthoryear{Palapanidis, Stergioulas  \&
  Lander}{Palapanidis et~al.}{2015}]{Palapanidis2015}
Palapanidis K.,  Stergioulas N.,   Lander S.~K.,  2015, \mnras, 452, 3246

\bibitem[\protect\citeauthoryear{{Passamonti} \& {Lander}}{{Passamonti} \&
  {Lander}}{2013}]{Passamonti2013}
{Passamonti} A.,  {Lander} S.~K.,  2013, \mn@doi [\mnras]
  {10.1093/mnras/sts372}, \href
  {http://adsabs.harvard.edu/abs/2013MNRAS.429..767P} {429, 767}

\bibitem[\protect\citeauthoryear{{Passamonti} \& {Lander}}{{Passamonti} \&
  {Lander}}{2014}]{Passamonti2014}
{Passamonti} A.,  {Lander} S.~K.,  2014, \mn@doi [\mnras]
  {10.1093/mnras/stt2134}, \href
  {http://adsabs.harvard.edu/abs/2014MNRAS.438..156P} {438, 156}

\bibitem[\protect\citeauthoryear{{Piro}}{{Piro}}{2005}]{Piro2005}
{Piro} A.~L.,  2005, \mn@doi [\apjl] {10.1086/499049}, \href
  {http://adsabs.harvard.edu/abs/2005ApJ...634L.153P} {634, L153}

\bibitem[\protect\citeauthoryear{{Prix} \& {Rieutord}}{{Prix} \&
  {Rieutord}}{2002}]{Prix2002}
{Prix} R.,  {Rieutord} M.,  2002, \mn@doi [\aap] {10.1051/0004-6361:20021049},
  \href {http://adsabs.harvard.edu/abs/2002A%26A...393..949P} {393, 949}

\bibitem[\protect\citeauthoryear{{Samuelsson} \& {Andersson}}{{Samuelsson} \&
  {Andersson}}{2007}]{Samuelsson2007}
{Samuelsson} L.,  {Andersson} N.,  2007, \mn@doi [\mnras]
  {10.1111/j.1365-2966.2006.11147.x}, \href
  {http://adsabs.harvard.edu/abs/2007MNRAS.374..256S} {374, 256}

\bibitem[\protect\citeauthoryear{{Samuelsson} \& {Andersson}}{{Samuelsson} \&
  {Andersson}}{2009}]{Samuelsson2009}
{Samuelsson} L.,  {Andersson} N.,  2009, \mn@doi [Classical and Quantum
  Gravity] {10.1088/0264-9381/26/15/155016}, \href
  {http://adsabs.harvard.edu/abs/2009CQGra..26o5016S} {26, 155016}

\bibitem[\protect\citeauthoryear{{Schulze} \& {Rijken}}{{Schulze} \&
  {Rijken}}{2011}]{Schulze2011}
{Schulze} H.-J.,  {Rijken} T.,  2011, \mn@doi [\prc]
  {10.1103/PhysRevC.84.035801}, \href
  {http://adsabs.harvard.edu/abs/2011PhRvC..84c5801S} {84, 035801}

\bibitem[\protect\citeauthoryear{{Shternin}, {Yakovlev}, {Heinke}, {Ho}  \&
  {Patnaude}}{{Shternin} et~al.}{2011}]{Shternin2011}
{Shternin} P.~S.,  {Yakovlev} D.~G.,  {Heinke} C.~O.,  {Ho} W.~C.~G.,
  {Patnaude} D.~J.,  2011, \mn@doi [\mnras] {10.1111/j.1745-3933.2011.01015.x},
  \href {http://adsabs.harvard.edu/abs/2011MNRAS.412L.108S} {412, L108}

\bibitem[\protect\citeauthoryear{{Sinha} \& {Sedrakian}}{{Sinha} \&
  {Sedrakian}}{2015}]{Sinha2015}
{Sinha} M.,  {Sedrakian} A.,  2015, \mn@doi [\prc]
  {10.1103/PhysRevC.91.035805}, \href
  {http://adsabs.harvard.edu/abs/2015PhRvC..91c5805S} {91, 035805}

\bibitem[\protect\citeauthoryear{{Sotani}}{{Sotani}}{2015}]{Sotani2015}
{Sotani} H.,  2015, Physical Review D, 92, 104024

\bibitem[\protect\citeauthoryear{{Sotani}, {Kokkotas}  \&
  {Stergioulas}}{{Sotani} et~al.}{2007}]{Sotani2007}
{Sotani} H.,  {Kokkotas} K.~D.,   {Stergioulas} N.,  2007, \mn@doi [\mnras]
  {10.1111/j.1365-2966.2006.11304.x}, \href
  {http://adsabs.harvard.edu/abs/2007MNRAS.375..261S} {375, 261}

\bibitem[\protect\citeauthoryear{{Sotani}, {Kokkotas}  \&
  {Stergioulas}}{{Sotani} et~al.}{2008}]{Sotani2008}
{Sotani} H.,  {Kokkotas} K.~D.,   {Stergioulas} N.,  2008, \mn@doi [\mnras]
  {10.1111/j.1745-3933.2007.00420.x}, \href
  {http://adsabs.harvard.edu/abs/2008MNRAS.385L...5S} {385, L5}

\bibitem[\protect\citeauthoryear{{Sotani}, {Nakazato}, {Iida}  \&
  {Oyamatsu}}{{Sotani} et~al.}{2013}]{Sotani2013}
{Sotani} H.,  {Nakazato} K.,  {Iida} K.,   {Oyamatsu} K.,  2013, \mn@doi
  [\mnras] {10.1093/mnrasl/sls006}, \href
  {http://adsabs.harvard.edu/abs/2013MNRAS.428L..21S} {428, L21}

\bibitem[\protect\citeauthoryear{{Sotani}, {Iida}  \& {Oyamatsu}}{{Sotani}
  et~al.}{2016}]{Sotani2016}
{Sotani} H.,  {Iida} K.,   {Oyamatsu} K.,  2016, \mn@doi [New Astronomy]
  {10.1016/j.newast.2015.08.003}, \href
  {http://adsabs.harvard.edu/abs/2016NewA...43...80S} {43, 80}

\bibitem[\protect\citeauthoryear{{Steiner} \& {Watts}}{{Steiner} \&
  {Watts}}{2009}]{Steiner2009}
{Steiner} A.~W.,  {Watts} A.~L.,  2009, \mn@doi [Physical Review Letters]
  {10.1103/PhysRevLett.103.181101}, \href
  {http://adsabs.harvard.edu/abs/2009PhRvL.103r1101S} {103, 181101}

\bibitem[\protect\citeauthoryear{{Steiner}, {Prakash}, {Lattimer}  \&
  {Ellis}}{{Steiner} et~al.}{2005}]{Steiner2005}
{Steiner} A.~W.,  {Prakash} M.,  {Lattimer} J.~M.,   {Ellis} P.~J.,  2005,
  \mn@doi [\physrep] {10.1016/j.physrep.2005.02.004}, \href
  {http://adsabs.harvard.edu/abs/2005PhR...411..325S} {411, 325}

\bibitem[\protect\citeauthoryear{{Stergioulas} \& {Friedman}}{{Stergioulas} \&
  {Friedman}}{1995}]{Stergioulas1995}
{Stergioulas} N.,  {Friedman} J.~L.,  1995, \mn@doi [\apj] {10.1086/175605},
  \href {http://adsabs.harvard.edu/abs/1995ApJ...444..306S} {444, 306}

\bibitem[\protect\citeauthoryear{{Strohmayer} \& {Watts}}{{Strohmayer} \&
  {Watts}}{2005}]{Strohmayer2005}
{Strohmayer} T.~E.,  {Watts} A.~L.,  2005, \mn@doi [\apjl] {10.1086/497911},
  \href {http://adsabs.harvard.edu/abs/2005ApJ...632L.111S} {632, L111}

\bibitem[\protect\citeauthoryear{{Strohmayer} \& {Watts}}{{Strohmayer} \&
  {Watts}}{2006}]{Strohmayer2006}
{Strohmayer} T.~E.,  {Watts} A.~L.,  2006, \mn@doi [\apj] {10.1086/508703},
  \href {http://adsabs.harvard.edu/abs/2006ApJ...653..593S} {653, 593}

\bibitem[\protect\citeauthoryear{{Thompson}, {Lyutikov}  \&
  {Kulkarni}}{{Thompson} et~al.}{2002}]{Thompson2002}
{Thompson} C.,  {Lyutikov} M.,   {Kulkarni} S.~R.,  2002, \mn@doi [\apj]
  {10.1086/340586}, \href {http://adsabs.harvard.edu/abs/2002ApJ...574..332T}
  {574, 332}

\bibitem[\protect\citeauthoryear{{Watts} \& {Strohmayer}}{{Watts} \&
  {Strohmayer}}{2006}]{Watts2006}
{Watts} A.~L.,  {Strohmayer} T.~E.,  2006, \mn@doi [\apjl] {10.1086/500735},
  \href {http://adsabs.harvard.edu/abs/2006ApJ...637L.117W} {637, L117}

\bibitem[\protect\citeauthoryear{{van Hoven} \& {Levin}}{{van Hoven} \&
  {Levin}}{2008}]{vanHoven2008}
{van Hoven} M.,  {Levin} Y.,  2008, \mn@doi [\mnras]
  {10.1111/j.1365-2966.2008.13881.x}, \href
  {http://adsabs.harvard.edu/abs/2008MNRAS.391..283V} {391, 283}

\bibitem[\protect\citeauthoryear{{van Hoven} \& {Levin}}{{van Hoven} \&
  {Levin}}{2011}]{vanHoven2011}
{van Hoven} M.,  {Levin} Y.,  2011, \mn@doi [\mnras]
  {10.1111/j.1365-2966.2010.17499.x}, \href
  {http://adsabs.harvard.edu/abs/2011MNRAS.410.1036V} {410, 1036}

\bibitem[\protect\citeauthoryear{{van Hoven} \& {Levin}}{{van Hoven} \&
  {Levin}}{2012}]{vanHoven2012}
{van Hoven} M.,  {Levin} Y.,  2012, \mn@doi [\mnras]
  {10.1111/j.1365-2966.2011.20177.x}, \href
  {http://adsabs.harvard.edu/abs/2012MNRAS.420.3035V} {420, 3035}

\makeatother
\end{thebibliography}

\end{document}